\documentclass{aa}  

\usepackage{graphicx}
\usepackage{txfonts}
\usepackage{lipsum}
\usepackage{subcaption}         
\usepackage{lscape}             
\usepackage{placeins}        
\usepackage{booktabs}
\usepackage{graphicx}  
\usepackage{makecell}
\usepackage{natbib}
\usepackage[colorlinks=true, urlcolor=green, citecolor=blue, linkcolor=blue, filecolor=red]{hyperref}

\begin{document}
   \title{Cluster-centric trends in bar size and pattern speed: the case of Abell 2199}

\author{Chandan Watts
          \inst{1,2}
          \and
          Sudhanshu Barway\inst{1,2}
          \and
          Mousumi Das\inst{1}
          \and
          Ewa L. Łokas\inst{3}
          }

   \institute{Indian Institute of Astrophysics,
              II Block, Koramangala, Bengaluru 560 034, India.\\
              \email{chandan@iiap.res.in} \and Pondicherry University, R.V. Nagar, Kalapet, 605014, Puducherry, India \and Nicolaus Copernicus Astronomical Center, Polish Academy of Sciences, Bartycka 18, 00-716 Warsaw, Poland }

   \date{}

  \abstract  
   {}
   {We investigate how the environment of a dynamically unrelaxed galaxy cluster influences the structure and dynamics of stellar bars. In particular, we examine cluster-centric variations in normalised bar size and bar pattern speed in Abell~2199.} 
   {Our analysis is based on 578 spectroscopically confirmed members of Abell~2199, including a master sample of 325 galaxies with homogeneous stellar mass and star formation rate measurements. We identify 39 barred galaxies and measure their structural properties using isophotal ellipse fitting and three-component (bulge+disc+bar) photometric decompositions. For 22 barred galaxies with MaNGA integral-field spectroscopy, we estimate bar pattern speeds using the Tremaine--Weinberg method, obtaining robust measurements for 12 galaxies. Stellar population age and projected specific angular momentum are analysed using $D4000_{R_{\mathrm{e}}}$ and $\lambda_{R_{\mathrm{e}}}$ from the MaNGA Pipe3D catalogue.}
   {Abell~2199 exhibits star formation--density and morphology--density relations despite its non-relaxed dynamical state. Early-type spiral (ETS) barred galaxies show systematic cluster-centric variations in normalised bar size, with relatively larger bars towards the cluster centre and smaller bars at intermediate radii. A corresponding variation in bar pattern speed with cluster-centric distance is also observed. These trends motivate a division at $\sim$0.5\,$R_{\mathrm{vir}}$, within which morphology-dependent environmental signatures become clearer, as barred galaxies in the inner region tend to host older stellar populations and lower projected angular momentum than those in the outskirts, with ETS+Bar galaxies retaining higher angular momentum compared to S0+Bar galaxies at comparable radii.}
   {}


   \maketitle
\nolinenumbers

\section{Introduction}
Galaxy evolution is governed by the interplay between internal processes and external environmental influences. Internal mechanisms, such as bar-driven secular evolution and active galactic nucleus (AGN) feedback, regulate the redistribution of gas and angular momentum, thereby shaping star formation, structural transformation, and kinematic properties. At the same time, global galaxy characteristics, including stellar mass and morphology, provide the framework within which these processes operate. A wide range of observational and theoretical studies have demonstrated that the combined action of these factors determines the evolutionary pathways of galaxies \citep{2021ApJ...921...60G, 2014ARA&A..52..291C, 2013ARA&A..51..511K, Kawinwanichakij_2017, Masters_2011, Zhuang_2021, Karim_2011, 10.1093/mnras/stt469, 2009ARA&A..47..159B}. 

The role of the environment is particularly significant, as galaxies evolve differently in the field, groups, and clusters, where distinct physical mechanisms dominate. Processes such as tidal interactions, ram-pressure stripping, harassment, and strangulation operate with varying efficiency depending on halo mass and local density \citep{1972ApJ...176....1G, 1998ApJ...495..139M, 1980ApJ...237..692L, 1972ApJ...178..623T, 1992ARA&A..30..705B, 2005Natur.433..604D, 2009MNRAS.399.2183N}. In cluster environments, galaxy properties further depend on cluster-centric distance, as galaxies located near the cluster core, at intermediate radii, and in the outskirts experience different interaction histories and intra-cluster medium (ICM) conditions \citep{2008MNRAS.391..585M, 2009ApJ...699.1595P, 2023ApJ...949...73V}.

Within clusters, the evolutionary impact of the environment also depends on the system's dynamical state \citep{2025A&A...699A.313A}. In dynamically young or merging clusters, galaxies are continuously accreted along large-scale filaments, and the presence of substructures can enhance tidal interactions and pre-processing \citep{2025A&A...702A.251V}. Such non-equilibrium conditions may modify galaxy morphology, star formation, and kinematic properties during infall towards the cluster centre, leading to evolutionary pathways that differ from those in dynamically relaxed systems.

Stellar bars are among the primary drivers of internal secular evolution in disc galaxies. By redistributing angular momentum and funnelling gas from the disc towards the central regions, bars can contribute to the build-up of central mass concentrations, regulate central star formation, and influence nuclear activity \citep{2014RvMP...86....1S,2019ApJ...872....5S,2018MNRAS.473.4731K}.

\begin{figure*}[htb!]
    \centering
    \includegraphics[width=0.8\linewidth]{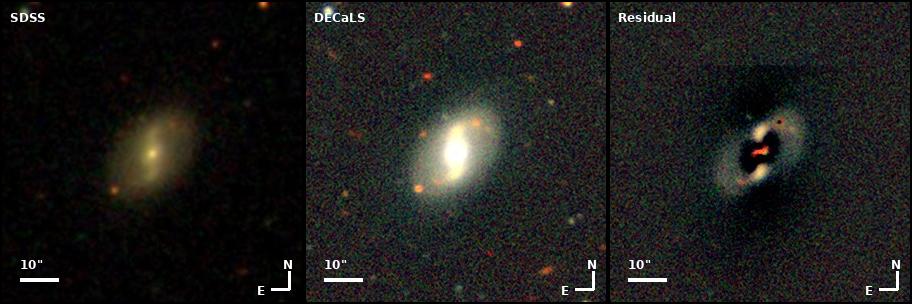}
    \caption{Example of an S0 barred galaxy (Bar ID 4). Colour-composite images ($300 \times 300$ pixels) are shown for comparison: SDSS (left), DECaLS (centre), and the corresponding DECaLS residual image (right). The DECaLS residual image is the best-fitting axisymmetric model, highlighting non-axisymmetric structures such as the bar.}
    \label{fig:morph}
\end{figure*}

Recent observational and theoretical studies have suggested that bar properties are not solely determined by internal dynamics but may also depend on the environment. In particular, variations in bar fraction, bar length, and bar pattern speed have been reported as a function of local density and cluster-centric distance \citep{2009A&A...497..713B,2014MNRAS.445.1339L,2016ApJ...826..227L,2018ApJ...857....6L,2025A&A...702A...7L,2026MNRAS.547ag175P,2026A&A...705A.115C}. These findings motivate a detailed investigation of barred galaxies within cluster environments, where the combined effects of tidal interactions, repeated encounters, and infall can be examined in a spatially resolved framework.

Because bar structure and bar dynamics encode information about angular-momentum redistribution and secular timescales, they provide a sensitive probe of how internal evolution couples to external environmental processes. A cluster-centric perspective is therefore essential for disentangling the role of environment in shaping barred galaxy evolution. 

Beyond bar structure and dynamics, the underlying stellar populations and angular-momentum distribution of barred galaxies provide additional constraints on their evolutionary pathways and environmental history \citep{10.1093/mnras/stae2252,10.1093/mnras/staf1760,Ansar_2024}. In particular, the connection between stellar angular momentum, morphology, and environment is highly relevant in cluster systems, where gravitational interactions and the cluster potential can perturb discs and alter their kinematic properties. Examining these quantities alongside bar properties, therefore, enables a more comprehensive view of how secular evolution and environmental processes operate in tandem.

Although stellar bars are frequently observed in disc galaxies, constructing robust samples of barred systems within galaxy clusters remains observationally challenging. Several studies have shown that the number of well-resolved barred galaxies decreases substantially when restrictions on morphology, inclination, stellar mass, and data quality are imposed \citep{2009A&A...497..713B,2014MNRAS.439.1749L,2023A&A...679A...5A}. Consequently, cluster-based investigations of barred galaxies often rely on modest sample sizes, particularly when detailed structural and dynamical measurements are required.

This study focuses on the galaxy cluster Abell~2199, a dynamically active system that has not yet reached full dynamical relaxation \citep{2009ChA&A..33....1Y}. Its non-equilibrium state makes it a particularly suitable laboratory for investigating how bar structure and dynamics respond to environmental conditions in clusters that are still assembling.

Abell~2199 ($z \sim 0.03$) is a rich cluster dominated by the central cD galaxy NGC~6166 and exhibits clear signatures of substructure and ongoing accretion, consistent with a non-virialised dynamical state \citep{bender.etal.201}. It forms part of a larger supercluster complex together with the nearby Abell~2197. Wide-area redshift surveys spanning $\sim$95\,deg$^{2}$ have mapped the large-scale infall of galaxies and groups towards the cluster centre \citep{2002AJ....124.1266R}, highlighting the continued growth of the system.

The structure of this paper is as follows. Section~2 describes the data and sample selection. Section~3 presents the analysis and results. Section~4 summarises our main conclusions. Throughout this work, we adopt a flat $\Lambda$CDM cosmology with $H_{0} = 70\,\mathrm{km\,s^{-1}\,Mpc^{-1}}$, $\Omega_{\mathrm{m}} = 0.286$, and $\Omega_{\Lambda} = 0.714$. At the redshift of Abell~2199, this corresponds to a physical scale of 0.608\,kpc\,arcsec$^{-1}$.

\section{Sample Selection and Data} \label{sec:sample}
In this section, we describe the imaging data sets and catalogues employed to construct the galaxy sample in Abell~2199, together with the derived parameters used in the subsequent analysis. Our goal is to assemble a homogeneous and internally consistent set of physical and morphological properties for cluster members spanning a range of cluster-centric distances.

\begin{figure*}[!htp]
 \centering
 \begin{subfigure}[b]{0.48\textwidth}
     \centering
     \includegraphics[width=\textwidth]{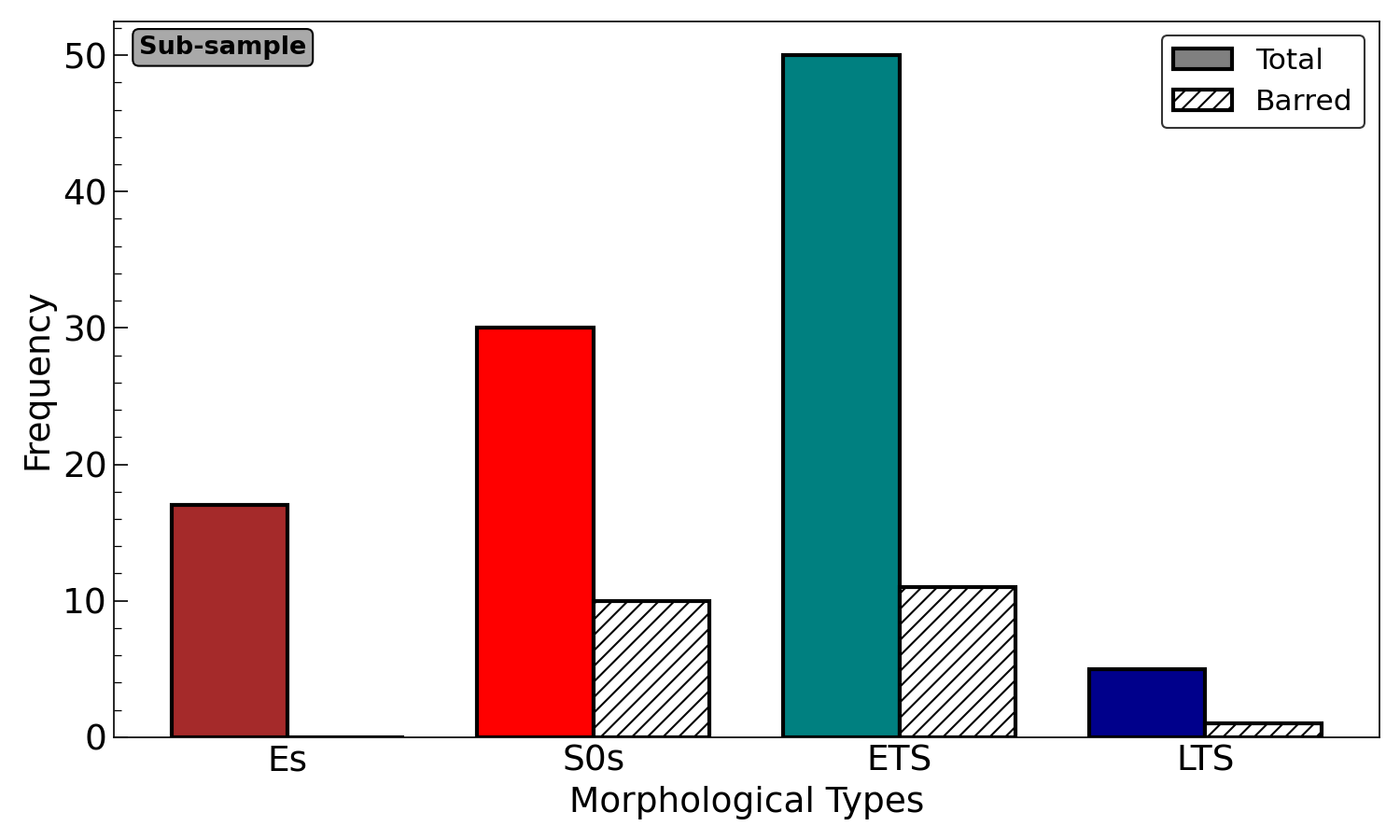}      
 \end{subfigure}
 \begin{subfigure}[b]{0.48\textwidth}
     \centering
     \includegraphics[width=\textwidth]{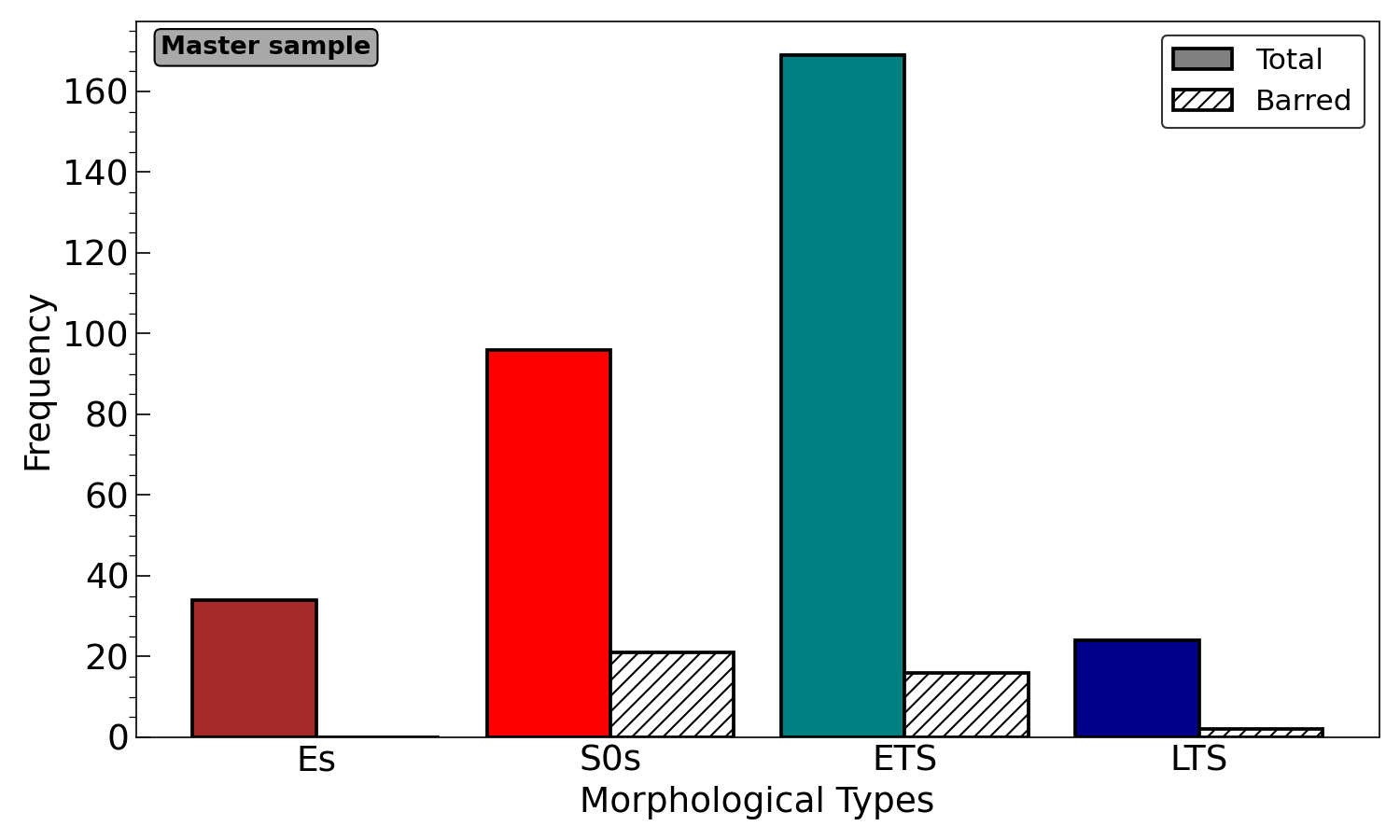}
 \end{subfigure}
 \caption{Morphological distribution of the galaxy sample in four broad classes: ellipticals (E), lenticulars (S0), early-type spirals (ETS), and late-type spirals (LTS). (left) MaNGA subsample (102 galaxies), showing the total number of galaxies in each morphological class and the subset identified as barred. (right) corresponding morphological distribution for the master sample of 325 galaxies.}
    \label{fig:sample_count}
\end{figure*}

\subsection{Sample Selection}
The galaxy sample analysed in this work is drawn from the catalogue of \citet{2017ApJ...842...88S}, based on a spectroscopic redshift survey of the nearby galaxy cluster Abell~2199. The survey provides measurements for 2275 galaxies in the cluster region (within radius $<$ 1$^\circ$, corresponding to $\sim$2.2 Mpc). Cluster membership is defined by applying a velocity cut of $7000 < cz < 11000$ km\,s$^{-1}$, resulting in a parent sample of 578 galaxies. This corresponds to $\pm$2.5$\sigma$, where $\sigma$ = 800 km\,s$^{-1}$ is the velocity dispersion of galaxies in the cluster \citep{2006MNRAS.367.1463L}. This cut is adopted to make sure that there are no interlopers from the background or foreground. In addition to membership information, the catalogue identifies galaxies associated with smaller groups embedded within the cluster environment.

To derive physical properties, in particular stellar mass and star formation rate (SFR), we cross-match the parent sample with the GALEX--SDSS--WISE Legacy Catalogue (GSWLC-X2; \citealt{2018ApJ...859...11S}). This catalogue provides parameters obtained from spectral energy distribution fitting based on ultraviolet, optical, and infrared data from \textit{GALEX}, SDSS, and \textit{WISE}. The cross-match yields a master sample of 325 galaxies, which forms the primary dataset for investigating cluster-centric variations in global galaxy properties. Of these, 314 galaxies also have measurements of local environmental density from \citet{2006MNRAS.373..469B}.

A complementary subsample is constructed by cross-matching the parent sample with the Mapping Nearby Galaxies at Apache Point Observatory (MaNGA) survey \citep{2015ApJ...798....7B}, part of SDSS-IV, and included in Data Release 17 (DR17). MaNGA provides spatially resolved spectroscopy for approximately $10^{4}$ galaxies over $\sim$2700\,deg$^{2}$, covering the wavelength range 360--1000\,nm with a median redshift of $z \sim 0.03$. This cross-match identifies 102 cluster members with integral-field spectroscopy. For these galaxies, we adopt measurements from the MaNGA Pipe3D value-added catalogue \citep{2022ApJS..262...36S}, including $D4000_{R_{\mathrm{e}}}$ and $\lambda_{R_{\mathrm{e}}}$ as tracers of stellar population age and projected specific angular momentum, respectively.

We therefore define three hierarchical data sets: (i) the parent sample of 578 spectroscopically confirmed cluster members, (ii) the master sample of 325 galaxies with homogeneous stellar mass and SFR measurements from GSWLC-X2, and (iii) the subsample of 102 galaxies with spatially resolved MaNGA spectroscopy. This tiered structure enables us to examine galaxy properties in Abell~2199 at both global and internal scales, and to explore how stellar populations and angular momentum vary with environment in a consistent framework.

\subsection{Imaging data}
This study makes use of imaging data from the Sloan Digital Sky Survey Data Release~7 (SDSS DR7; \citealt{2009ApJS..182..543A}) and the Dark Energy Camera Legacy Survey Data Release~10 (DECaLS DR10; \citealt{2019AJ....157..168D}). SDSS SkyServer images, Legacy Survey colour composites, and their corresponding residual maps are employed for visual inspection and morphological classification.

SDSS provides imaging in five broad optical bands ($u$, $g$, $r$, $i$, and $z$) obtained with the 2.5\,m telescope at Apache Point Observatory. The typical point spread function (PSF) in the $r$ band is $\sim$1.32\arcsec, and the pixel scale is 0.396\arcsec\,pixel$^{-1}$. The survey reaches 95\% completeness limits of approximately 22.0, 22.2, 22.2, 21.3, and 20.5\,mag (AB) in the $u$, $g$, $r$, $i$, and $z$ bands, respectively.

The DECaLS DR10 images are obtained from the DECam Legacy Survey \citep{2019AJ....157..168D}. The survey covers $\sim$19\,000\,deg$^{2}$ at high Galactic latitudes, primarily in the northern and southern Galactic caps. Imaging is conducted with the DECam instrument on the Blanco 4\,m telescope in the $g$, $r$, and $z$ bands over the wavelength range 400--1000\,nm. The survey reaches 5$\sigma$ depths of 23.72, 23.27, and 22.22\,mag (AB) in the $g$, $r$, and $z$ bands, respectively, for an exponential galaxy profile with a half-light radius of 0.45\arcsec. DECaLS DR10 provides co-added images with a pixel scale of 0.262\arcsec\,pixel$^{-1}$.

\begin{figure}[ht!]
    \centering
    \includegraphics[width=\linewidth]{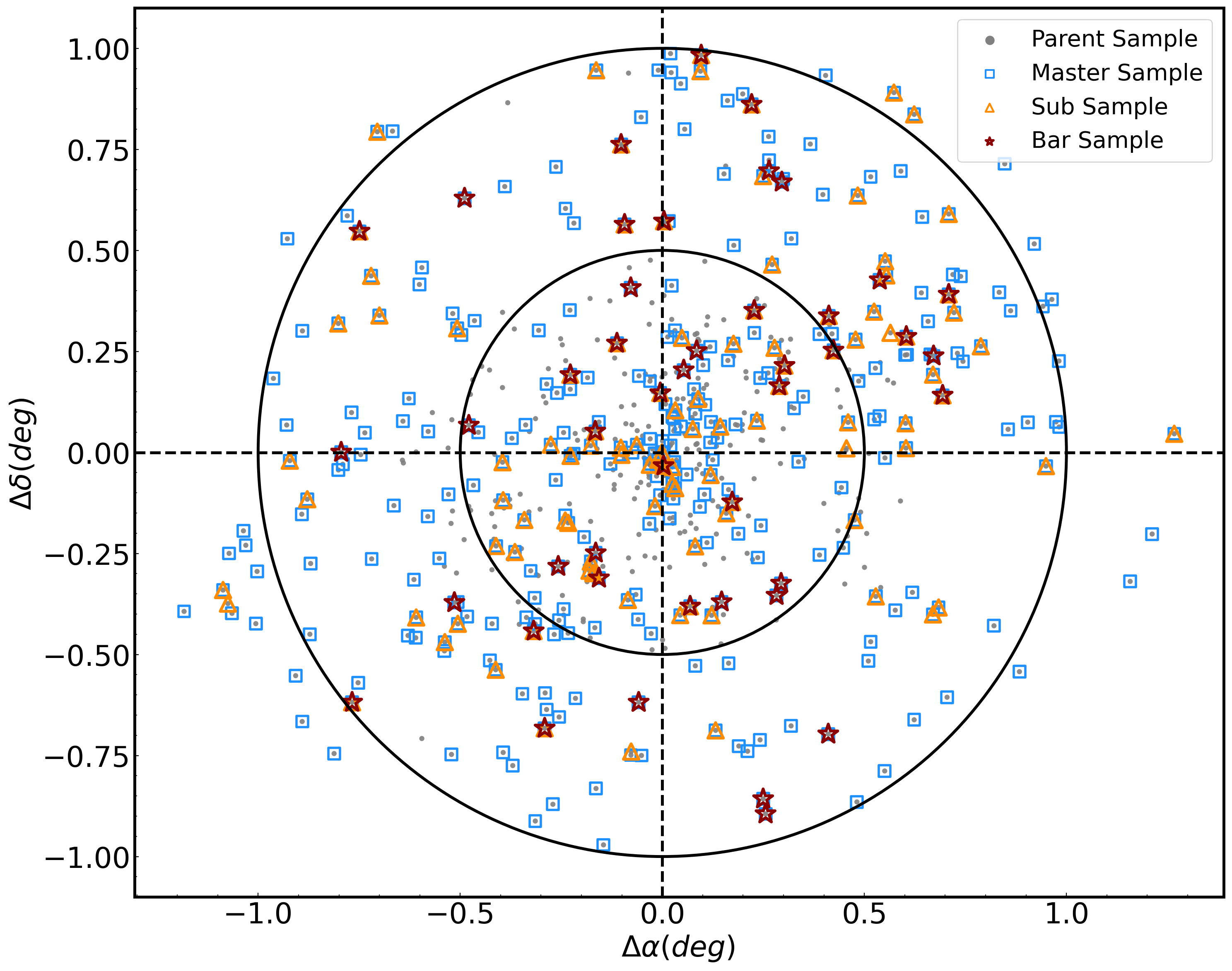}
    \caption{Projected spatial distribution of galaxies in the Abell~2199 cluster. Grey points represent the full parent sample of 578 spectroscopically confirmed members. Blue open squares indicate the master sample of 325 galaxies with stellar mass and sSFR measurements. Orange open triangles denote the MaNGA subsample (102 galaxies), and red open stars mark the barred galaxies (39 galaxies). The dashed lines indicate the adopted cluster centre (NGC 6166), and the concentric circles correspond to projected radii of $\sim$1.1 and 2.2\,Mpc.}
    \label{fig:cluster}
\end{figure}

\subsection{Visual Morphology Classification}
Morphological information is not available for all galaxies in the sample. For the subsample of 102 MaNGA galaxies, classifications are adopted from the MaNGA Visual Morphology Catalogue \citep{2022MNRAS.512.2222V}, which is based on SDSS and DESI Legacy Survey imaging. We group galaxies into four broad morphological classes: ellipticals (E), lenticulars (S0: S0, S0/a), early-type spirals (ETS: Sa, Sab, Sb, Sbc), and late-type spirals (LTS: Sc, Scd, Sd, Sdm, Sm, Im, S). Morphological classifications are therefore available for all 102 galaxies in the MaNGA subsample (Fig.~\ref{fig:sample_count}).

To obtain a homogeneous classification for the full master sample of 325 galaxies, we performed a visual inspection of SDSS, DECaLS, and corresponding residual images (Fig.~\ref{fig:morph}). To ensure consistency, we first calibrated our classification scheme using the 102 galaxies with existing MaNGA-based morphologies and subsequently applied the same criteria to the remaining galaxies. The same four broad morphological categories (E, S0, ETS, and LTS) are adopted throughout. Each galaxy was classified independently three times, and the final assignment was cross-checked against the probabilistic classifications of \citet{2011A&A...525A.157H} for 313 galaxies. In cases of disagreement—primarily between E and S0, or between ETS and LTS, we retained the visual classification. This procedure provides a uniform morphological data set for the entire master sample (Fig.~\ref{fig:sample_count}).

Barred galaxies were identified within this morphological framework. A total of 62 galaxies in the master sample are initially flagged as barred, together with three additional barred galaxies from the parent sample that are not included in the master sample, yielding 65 candidate barred galaxies in the cluster. To ensure that these 65 galaxies truly host bars and are not features caused by poor residuals or artifacts that might mimic bars, we examined the $r-$ and $i-$band images. Bar presence was then verified using these images. Of these, 39 galaxies allowed reliable bar modelling, with bar parameters derived using the ellipticity method (see Sect.~\ref{sec:bar}). The remaining 26 galaxies were excluded owing to high inclinations, the absence of a clear ellipticity maximum, or the inability to obtain a stable multi-component fit with \textsc{GALFIT} \citep{2002AJ....124..266P,2010AJ....139.2097P}. We therefore define a final bar sample consisting of these 39 galaxies. However, stellar mass and SFR information are not available for one galaxy (Bar\_ID = 5), as it is not part of the master sample. The spatial distribution of the parent, master, subsample, and bar samples within Abell~2199 is shown in Fig.~\ref{fig:cluster}.

In the analysis that follows, the master sample is used to characterise the global cluster properties, while the MaNGA subsample provides spatially resolved information on stellar populations and angular momentum. The bar sample serves as the basis for our investigation of bar structure and dynamics within the cluster environment.

\begin{figure}[ht!]
    \centering
    \includegraphics[width=\columnwidth]{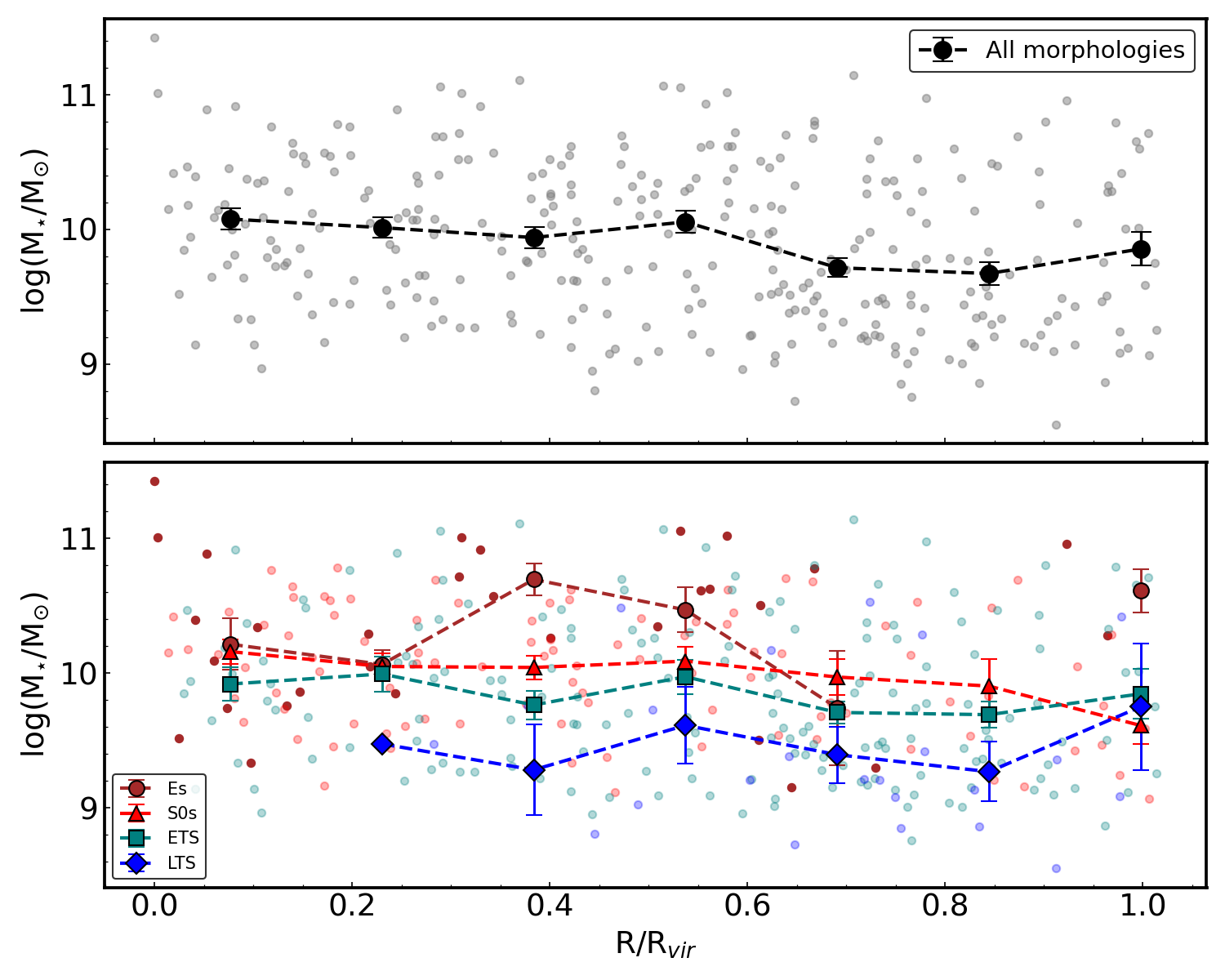}
    \caption{Stellar mass as a function of projected cluster-centric distance for (top) all morphologies combined, and (bottom) different morphological classes: ellipticals (E; brown), lenticulars (S0; red), early-type spirals (ETS; teal), and late-type spirals (LTS; blue). Scattered points represent individual galaxies, and points with dashed lines denote bin-averaged values computed using a radial bin size of 0.35 Mpc.}
    \label{fig:mass}
\end{figure}

\section{Results and Discussion}
Using the samples defined in Sect.~\ref{sec:sample}, we investigate cluster-centric variations in a range of galaxy properties, including stellar mass, specific star formation rate (sSFR), morphology, bar structure and dynamics, projected specific angular momentum, and stellar population age. 

\begin{figure*}[htb!]
 % \hspace*{-1.0cm}
     \centering
     \begin{subfigure}[b]{0.45\textwidth}
         \centering
         \includegraphics[width=\textwidth]{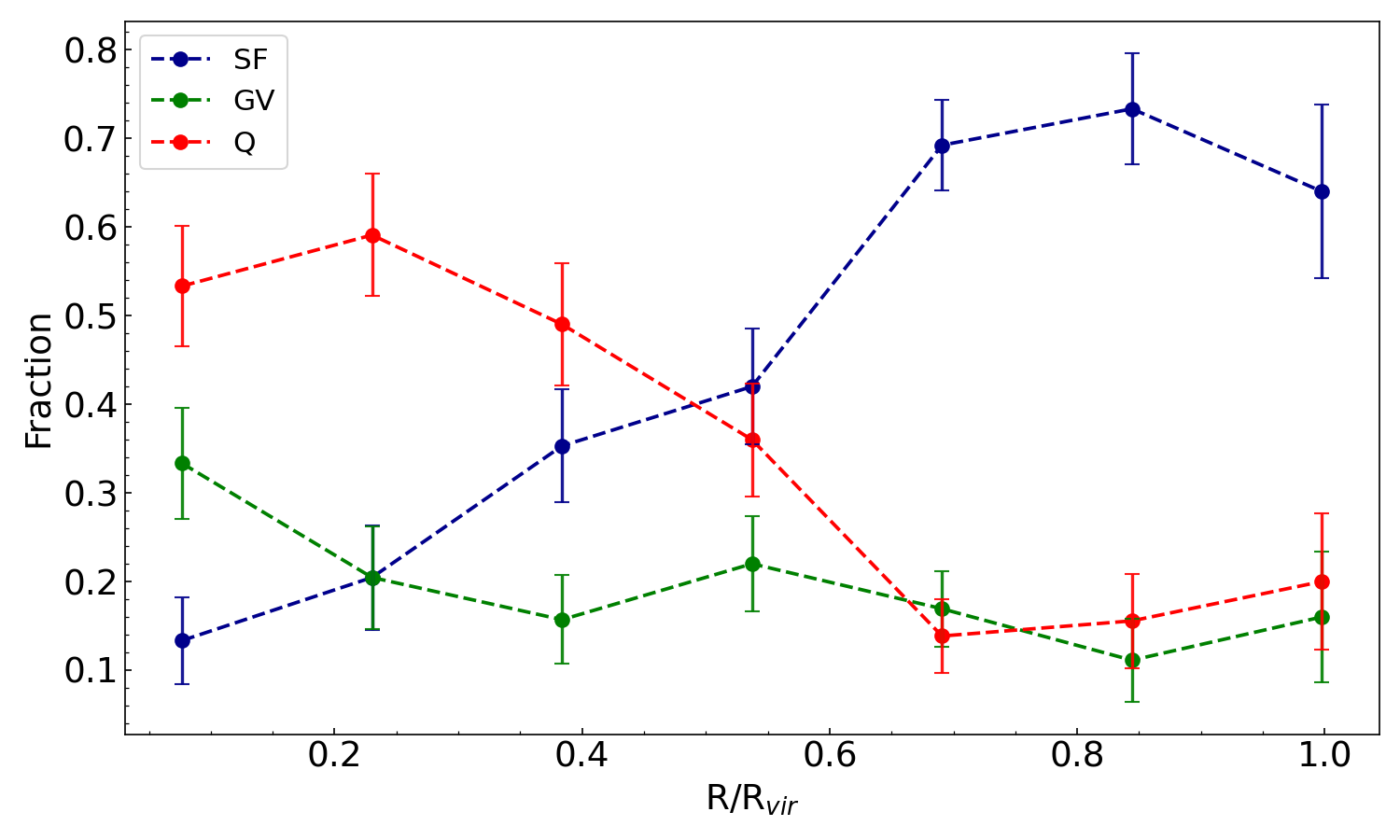}
     \end{subfigure}
     \begin{subfigure}[b]{0.45\textwidth}
         \centering
         \includegraphics[width=\textwidth]{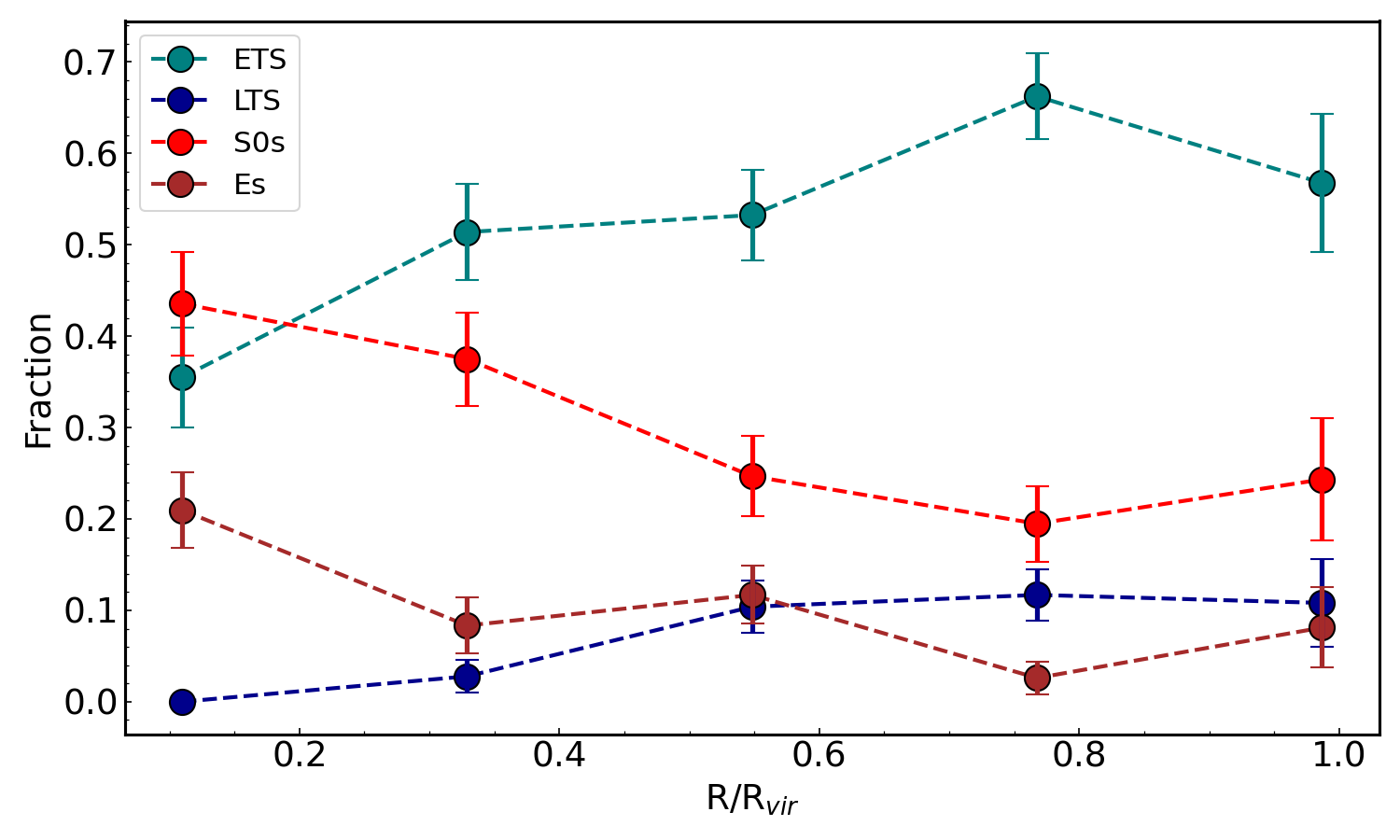}
     \end{subfigure}
     \caption{Cluster-centric trends for the master sample. (left) Fraction of star-forming (SF; blue), green-valley (GV; green), and quenched (Q; red) galaxies as a function of projected cluster-centric distance, using a radial bin size of 0.35\,Mpc. (right) Fraction of galaxies in different morphological classes, ellipticals (E; brown), lenticulars (S0; red), early-type spirals (ETS; teal), and late-type spirals (LTS; blue), as a function of projected cluster-centric distance, using a bin size of 0.5\,Mpc. Fractions are computed relative to the total number of galaxies in each radial bin. Error bars represent bootstrap uncertainties.}
      \label{fig:moph_var}
\end{figure*}

As an initial consistency check, we examine the distribution of stellar mass in the master sample as a function of projected cluster-centric distance. All trends in this work are based on projected cluster-centric distances. As a result, projection effects can introduce scatter and dilute intrinsic gradients, particularly in a non-relaxed cluster like Abell 2199, where recovering true three-dimensional distances is challenging. The observed trends should therefore be interpreted with appropriate caution. All cluster-centric distances are normalised by the virial radius (R$_{vir}$).  We consider the virial radius of Abell~2199, $R_{\mathrm{vir}} = 1.6\,h^{-1}$\,Mpc \citep{2002AJ....124.1266R}, corresponding to $R_{\mathrm{vir}} \simeq 2.28$\,Mpc for $h=0.7$. This value is similar to that obtained by \citet{2006MNRAS.367.1463L} from modeling.

We find that stellar mass is distributed across the full radial extent of the cluster without any statistically significant trend (Fig.~\ref{fig:mass}). A weak decrease in stellar mass is visible in the outermost radial bins; however, this behaviour is driven primarily by E galaxies, while the other morphological classes do not show a clear variation with cluster-centric distance.

\subsection{SF-Density Relation} \label{sec:sf}
\citet{2017ApJ...842...88S} provides information on whether cluster members are associated with galaxy groups or are isolated systems (i.e., not assigned to any group but still within the cluster potential). Within the master sample of 325 galaxies, 225 are classified as group members and 100 as isolated.

We first examine the distribution of local environmental density, adopting the local environmental density parameter $\Sigma$ from \citet{2006MNRAS.373..469B}. Galaxies are divided into three regimes: low density ($\log \Sigma\,(\mathrm{Mpc^{-2}}) < -0.5$), intermediate density ($-0.5 < \log \Sigma\,(\mathrm{Mpc^{-2}}) < 0.5$), and high density ($\log \Sigma\,(\mathrm{Mpc^{-2}}) > 0.5$). Isolated galaxies are predominantly found near the boundary between the intermediate- and high-density regimes, whereas group members are largely concentrated in the high-density environment (Fig.~\ref{fig:sample}).

We next consider the specific star formation rate (sSFR), adopting the classification scheme of \citet{2014SerAJ.189....1S}: star-forming ($\log(\mathrm{sSFR}) \geq -10.8$), green valley ($-11.8 < \log(\mathrm{sSFR}) < -10.8$), and quenched ($\log(\mathrm{sSFR}) \leq -11.8$). The master sample is dominated by star-forming galaxies. Isolated galaxies are primarily located in the star-forming regime, whereas group members show a bimodal distribution with significant fractions in both the star-forming and quenched categories (Fig.~\ref{fig:sample}). Only five isolated galaxies lie within a projected radius of 1.1\,Mpc from the cluster centre.

To assess the radial dependence of star formation activity, we compute the fractions of star-forming, green valley, and quenched galaxies as a function of projected cluster-centric distance (Fig.~\ref{fig:moph_var}). The fraction of quenched galaxies increases towards the cluster centre, while star-forming galaxies dominate beyond \textbf{$\sim$0.5}, indicating a transition in galaxy populations with radius. The green valley fraction is elevated in the central region and remains comparatively stable across the cluster.

\subsection{Morphology-Density Relation}

The morphology--density relation in rich clusters was first established observationally by \citet{1980ApJ...236..351D}, and has since been reproduced in cosmological simulations, including the EAGLE simulations \citep{2023MNRAS.518.5260P}. We therefore examine whether a similar environmental dependence of morphology is present in Abell~2199, despite its dynamically unrelaxed state.

Using the master sample, we compute the fractions of E, S0, ETS, and LTS as a function of projected cluster-centric distance (Fig.~\ref{fig:moph_var}). The central regions are dominated by early-type galaxies, with S0 galaxies constituting the largest fraction, followed by E. Moving towards larger cluster-centric distances, the fraction of S0 galaxies decreases, while the relative contribution of ETS galaxies increases. LTS remain a minor component throughout.

Although the gradients are weaker than those typically observed in dynamically relaxed clusters, the overall trend is consistent with the classical morphology-density relation. The presence of this relation in Abell\,2199 indicates that environmental effects are already shaping the morphological mix, even in a system that has not yet reached dynamical equilibrium. 

This morphology-dependent variation with cluster-centric distance is particularly relevant for the analysis of barred galaxies that follows, as bar properties are expected to depend on host morphology. Establishing the underlying morphological distribution, therefore, provides an essential framework for interpreting environmental trends in bar structure and dynamics.

\begin{figure*}[htb!]
     \centering
     \begin{subfigure}[b]{0.45\textwidth}
         \centering
         \includegraphics[width=\textwidth]{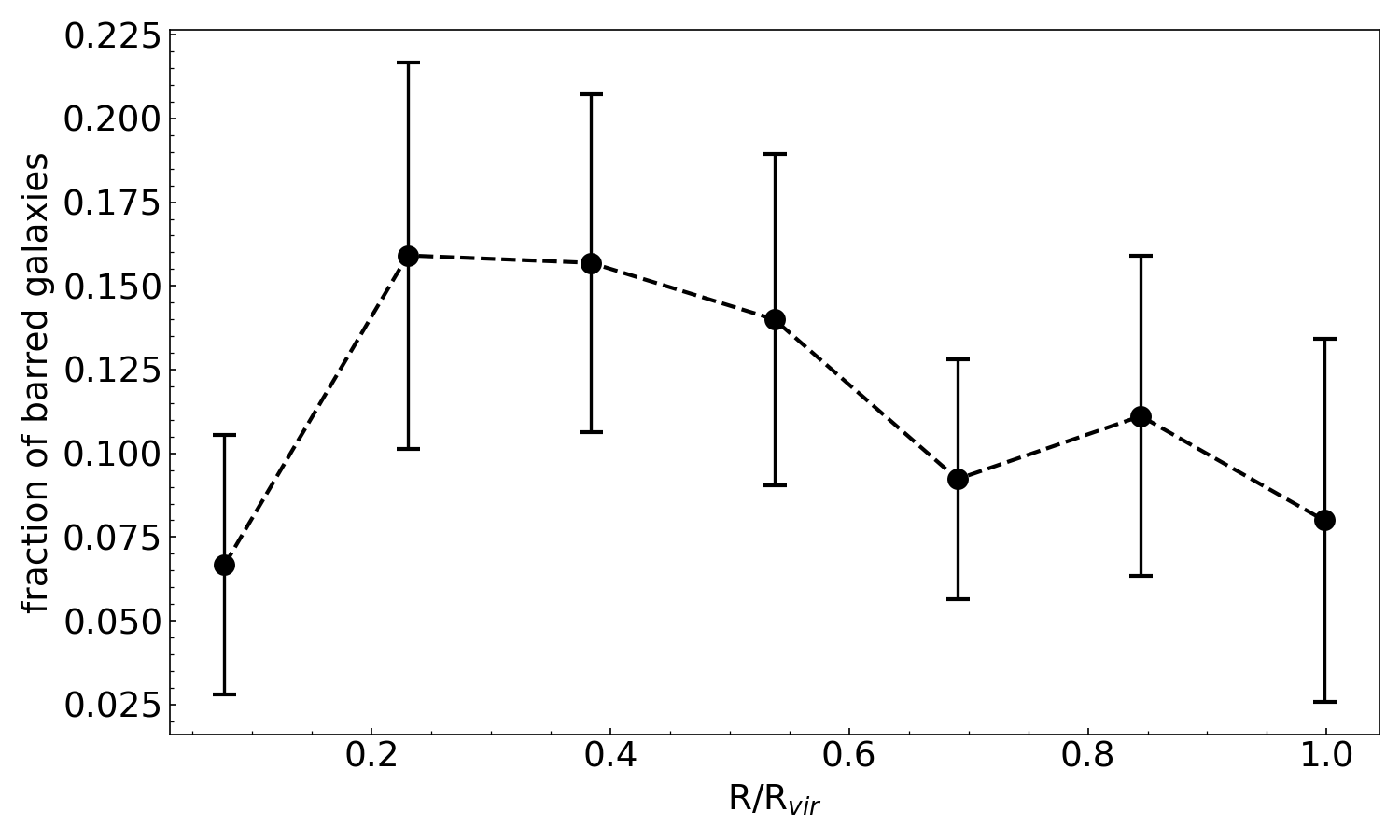}
     \end{subfigure}
     \begin{subfigure}[b]{0.45\textwidth}
         \centering
         \includegraphics[width=\textwidth]{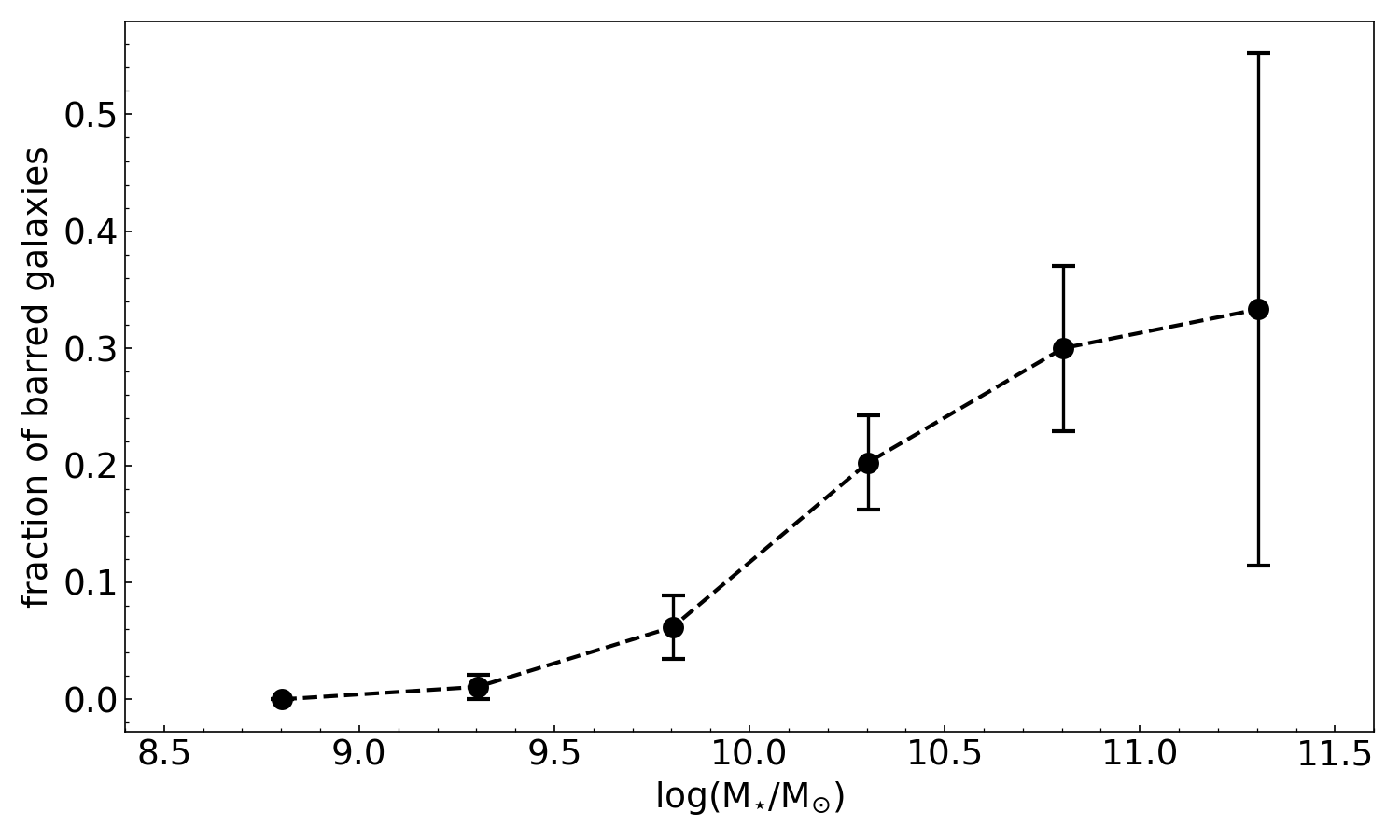}
     \end{subfigure}
     \caption{(left) Fraction of barred galaxies as a function of projected cluster-centric distance, using a bin size of 0.35\,Mpc. Fractions are computed relative to the total number of galaxies in the master sample within each radial bin. The number of barred galaxies in each radial bin is 3/42, 7/37, 8/43, 7/43, 6/59, 5/40, and 2/23 (Barred/Total). (right) Fraction of barred galaxies as a function of stellar mass, using a bin size of 0.5\,M$_{\odot}$. Fractions are computed relative to the total number of galaxies in the master sample within each mass bin. The number of barred galaxies in each mass bin is 0/15, 1/94, 5/81, 18/89, 12/40, and 2/6 (Barred/Total). Error bars represent bootstrap uncertainties.}
    \label{fig:bar_fraction_cluster_center}
\end{figure*}

\begin{figure*}[htb!]
     \hspace*{-1.0cm}
     \centering   
     \begin{subfigure}[b]{0.45\textwidth}
         \centering
         \includegraphics[width=\textwidth]{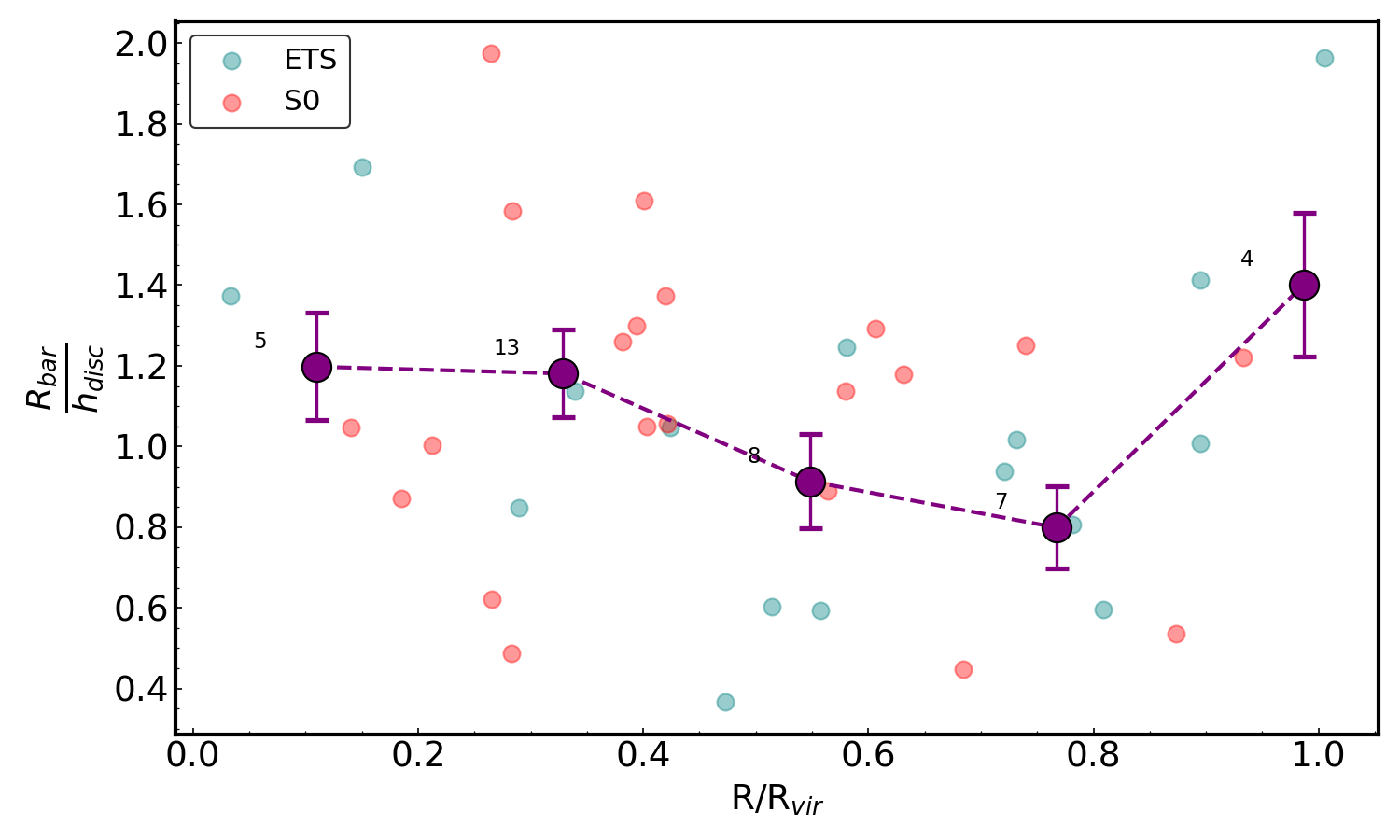}
     \end{subfigure}
     \begin{subfigure}[b]{0.45\textwidth}
         \centering
         \includegraphics[width=\textwidth]{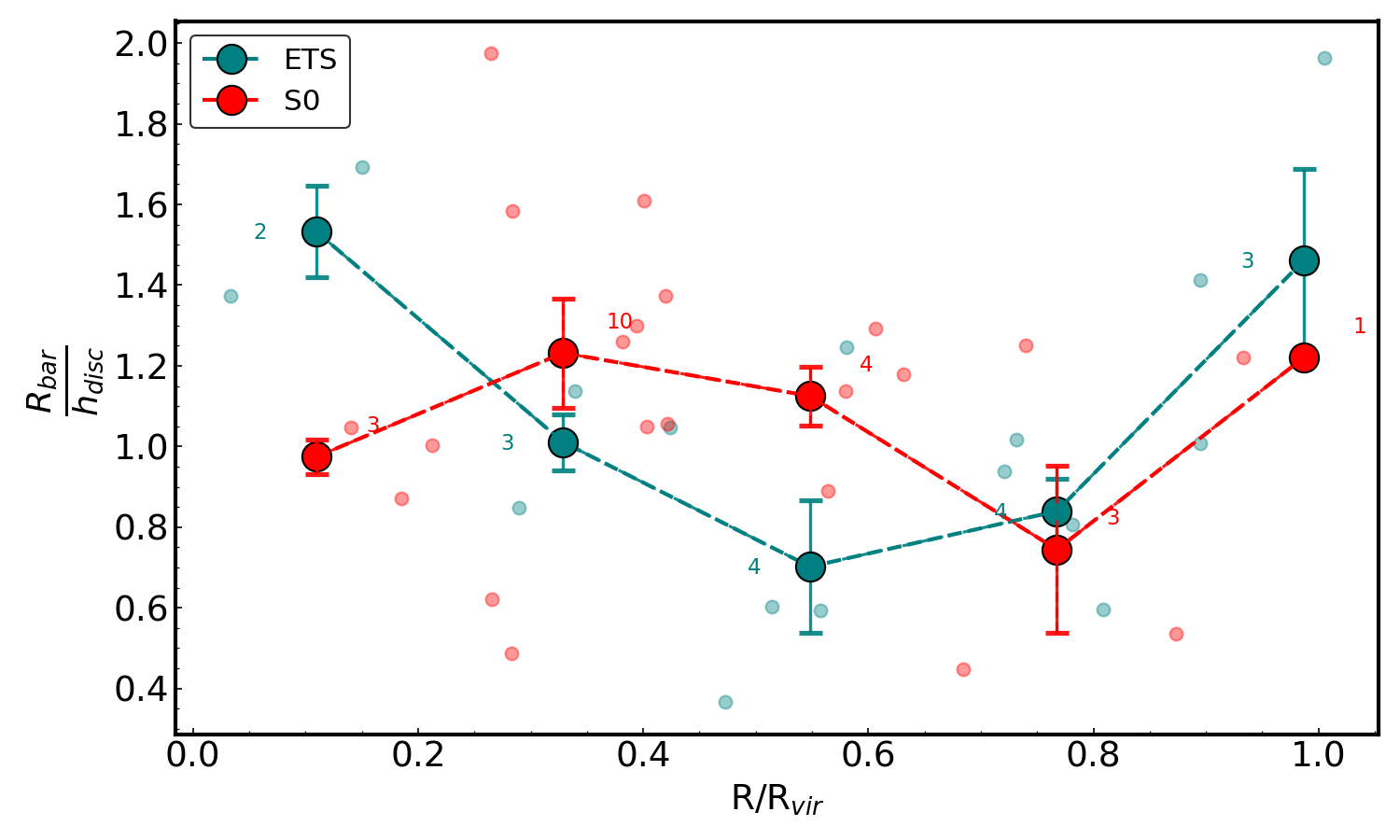}
     \end{subfigure}
     \caption{Cluster-centric trends for the barred galaxy sample. (left) Normalised bar radius, $R_{\mathrm{bar}}/h_{\mathrm{disc}}$, as a function of projected cluster-centric distance for all barred galaxies. (right) Same quantity  as a function of projected cluster-centric distance for ETS+Bar (teal) and S0+Bar (red) galaxies. Here, $h_{\mathrm{ disc}}$ is the disc scale length, derived from disc effective radius obtained from photometric decomposition. Points represent individual galaxies, and solid symbols denote bin-averaged values computed using a radial bin size of 0.5\,Mpc. Error bars indicate the standard deviation within each bin. The numbers shown on the solid symbol indicate the number of galaxies per bin.}
        \label{fig:bar_var_disc}
\end{figure*}

\subsection{Barred Galaxies} \label{sec:bar}
From the morphological classification described above, we identify a final barred galaxy sample of 39 galaxies within Abell~2199, comprising 16 ETS+Bar, 21 S0+Bar, and 2 LTS+Bar galaxies. We examined the fraction of barred galaxies as a function of projected cluster-centric distance, as shown in the left panel of Figure~\ref{fig:bar_fraction_cluster_center}. The bar fraction is lowest near the cluster centre, increases at intermediate radii, and then declines toward larger distances. This suggests that the role of cluster environment in inducing or destroying bars may be more complicated than envisioned in earlier work. Some previous studies found the bar fraction to increase toward the cluster centre \citep{Thompson1981, Andersen1996, 2009A&A...497..713B, 2014MNRAS.439.1749L} while other ones found little or no significant dependence of the bar fraction on cluster-centric distance \citep{2010ApJ...711L..61M, Cervantes2015, 2016ApJ...826..227L, Tawfeek2022, 2023A&A...679A...5A}. Taking our result at face value, one could speculate that the tidal forces are most efficient in inducing bars in the inner regions of the cluster, but not at the very center, where they are more likely to destroy them. 

We also examined the barred galaxy fraction in different stellar mass bins, as shown in the right panel of Figure~\ref{fig:bar_fraction_cluster_center}. We find a similar trend to that reported by \citet{2010ApJ...711L..61M, 2023A&A...678A..54M} for the Coma, Virgo, and SMACS0723 galaxy clusters. Given the small number of LTS+Bar, our analysis primarily focuses on the ETS+Bar and S0+Bar populations, which together constitute the dominant barred morphologies in the cluster environment.

This section presents the structural and dynamical properties of these barred galaxies as a function of cluster-centric distance. In particular, we examine variations in the normalised bar radius and bar pattern speed, which provide complementary diagnostics of bar evolution in different environments. Although the sample size is modest, the data suggest possible trends, especially among ETS+Bar galaxies, that are explored in the following subsections. Where appropriate, we interpret these results cautiously, given the limited sample size.

\subsubsection{Bar Length}
Bar lengths are measured using SDSS $r$-band images. Although several techniques are available for estimating bar extent, we adopt the isophotal ellipse-fitting method, which is widely used in the literature and provides consistent estimates of bar length \citep{2009A&A...495..491A,2019MNRAS.482.1733G}.

In this method, the galaxy surface-brightness distribution is fitted with a series of concentric elliptical isophotes. The ellipticity typically increases along the bar, reaches a maximum close to the bar end, and subsequently declines as the disc component becomes dominant. We therefore adopt the semi-major axis corresponding to the maximum ellipticity as the projected bar radius. An illustrative example (Bar ID 4) is shown in Fig.~\ref{fig:ellip}. To estimate the uncertainty in this measurement, we follow \citet{2007ApJ...657..790M} and consider the range of semi-major axis values satisfying $\epsilon \geq (\epsilon_{\mathrm{max}} - 0.01)$, which defines the extent of the ellipticity peak. Following this procedure, bar lengths and their uncertainties are measured for all 39 galaxies in the barred sample (Table~\ref{tab:bar_props}).

The measured values correspond to projected bar radii. To account for projection effects, we deproject the bar length using the one-dimensional approximation of \citet{2007MNRAS.381..943G}:

\begin{equation}
R_{\mathrm{bar,deproj}} = R_{\mathrm{bar,proj}}
\left( \sin^{2}\alpha \,\sec^{2} i + \cos^{2}\alpha \right)^{1/2},
\end{equation}
where $\alpha$ is the position-angle offset between the bar and the galaxy's major axis, and $i$ is the inclination angle of the disc. Galaxy position angles and inclinations are adopted from the NASA--Sloan Atlas (NSA; version $V\_1\_0\_1$; \citealt{2011AJ....142...31B}). For one galaxy (Bar ID 28) lacking NSA measurements, we determine the position angle and inclination from our isophotal analysis at the radius enclosing 90\% of the total flux, consistent with the NSA definition. The projected and deprojected bar properties are listed in Table~\ref{tab:bar_props}.

We note that the ellipticity-based definition of bar length may be influenced by projection effects, bulge dominance, or the presence of spiral arms, all of which can shift the position of the maximum ellipticity (e.g. \citealt{2007MNRAS.381..943G, 2015ApJ...799..226E}). In particular, in galaxies with prominent bulges or lenses, the ellipticity peak may not coincide exactly with the dynamical bar end. Moreover, different bar-length definitions can yield systematically different values \citep{2008AJ....135...20E}. Nevertheless, the ellipticity method provides a consistent and widely adopted estimate of relative bar size, and is therefore appropriate for investigating cluster-centric trends within a homogeneous framework \citep{das.etal.2003}.

\begin{table*}[ht!]
\centering
\caption{Measured bar parameters for the barred galaxy sample.}
\label{tab:bar_props}
\resizebox{1.0\textwidth}{!}{%
\begin{tabular}{@{}llllllllllll@{}}
\toprule \toprule
Bar ID &
\makecell{RA \\ (deg)} &
\makecell{Dec \\ (deg)} &
Morphology &
\makecell{$R_{\mathrm{bar}}^{\mathrm{proj}}$ \\ (arcsec)} &
\makecell{$R_{\mathrm{bar}}^{\mathrm{deproj}}$ \\ (kpc)} &
\makecell{PA$_{\mathrm{bar}}$ \\ (deg)} &
\makecell{$R_{\mathrm{bar}}/R_{\mathrm{50}}$} &
\makecell{$R_{\mathrm{bar}}/R_{\mathrm{90}}$} &
\makecell{PA$_{\mathrm{gal}}$ \\ (deg)} &
\makecell{Inclination \\ (deg)} &
\makecell{SFR Class \\ (SF/GV/Q)} 
\\ \midrule
1 & 247.4562 & 40.2208 & ETS + Bar & 5.22$\pm$0.26 & 3.30$\pm$0.16 & 129.00$\pm$1.23 & 1.08 & 0.41 & 88.45 & 24.11 & Q \\
2 & 247.8315 & 39.7905 & ETS + Bar & 9.71$\pm$0.43 & 5.92$\pm$0.26 & 96.73$\pm$0.61 & 1.52 & 0.61 & 100.89 & 45.22 & Q \\
3 & 247.5724 & 39.8894 & ETS + Bar & 2.14$\pm$0.11 & 1.30$\pm$0.07 & 56.51$\pm$0.59 & 0.37 & 0.13 & 59.48 & 43.29 & Q \\
4 & 246.6816 & 39.6187 & S0 + Bar & 6.39$\pm$0.32 & 4.00$\pm$0.20 & 175.32$\pm$0.83 & 1.20 & 0.44 & 161.91 & 46.06 & Q \\
5 & 247.4627 & 39.7665 & LTS + Bar & 14.19$\pm$0.42 & 8.76$\pm$0.26 & 5.92$\pm$0.73 & 0.78 & 0.34 & 173.29 & 39.54 & $-$ \\
6 & 247.4543 & 39.227 & S0 + Bar & 3.89$\pm$0.19 & 2.37$\pm$0.12 & 82.38$\pm$0.70 & 0.93 & 0.38 & 88.12 & 28.77 & Q \\
7 & 247.2453 & 39.8031 & S0 + Bar & 2.62$\pm$0.13 & 1.64$\pm$0.08 & 121.59$\pm$0.78 & 0.51 & 0.18 & 107.52 & 42.84 & Q \\
8 & 247.3878 & 39.9030 & S0 + Bar & 5.66$\pm$0.28 & 3.44$\pm$0.17 & 160.42$\pm$0.35 & 1.91 & 0.66 & 158.03 & 47.28 & Q \\
9 & 247.8691 & 39.9428 & S0 + Bar & 3.23$\pm$0.08 & 1.98$\pm$0.05 & 174.18$\pm$1.56 & 0.62 & 0.22 & 10.77 & 24.66 & SF \\
10 & 247.7642 & 39.8385 & ETS + Bar & 6.10$\pm$0.15 & 3.97$\pm$0.10 & 151.04$\pm$1.14 & 0.61 & 0.23 & 120.05 & 36.39 & GV \\
11 & 247.4099 & 38.6942 & ETS + Bar & 2.86$\pm$0.07 & 1.75$\pm$0.04 & 39.37$\pm$1.05 & 0.70 & 0.31 & 48.11 & 34.85 & SF \\
12 & 246.8693 & 38.8691 & ETS + Bar & 8.45$\pm$0.21 & 6.05$\pm$0.15 & 5.04$\pm$0.65 & 0.99 & 0.32 & 102.84 & 32.07 & Q \\
13 & 247.0823 & 39.9591 & S0 + Bar & 3.72$\pm$0.09 & 2.27$\pm$0.06 & 59.39$\pm$0.75 & 1.25 & 0.44 & 68.16 & 30.44 & Q \\
14 & 247.4245 & 40.2482 & S0 + Bar & 5.41$\pm$0.14 & 3.29$\pm$0.08 & 116.46$\pm$1.11 & 1.64 & 0.50 & 118.32 & 36.19 & Q \\
15 & 247.4159 & 38.6572 & S0 + Bar & 4.08$\pm$0.10 & 2.50$\pm$0.06 & 85.65$\pm$0.65 & 1.21 & 0.41 & 91.12 & 48.53 & Q \\
16 & 247.4430 & 39.1978 & ETS + Bar & 4.65$\pm$0.12 & 2.84$\pm$0.07 & 20.57$\pm$0.42 & 0.85 & 0.31 & 15.65 & 52.27 & SF \\
17 & 246.4111 & 40.0988 & ETS + Bar & 3.81$\pm$0.10 & 2.48$\pm$0.06 & 69.01$\pm$0.60 & 0.96 & 0.30 & 94.14 & 41.49 & GV \\
18 & 247.0036 & 39.2407 & ETS + Bar & 3.87$\pm$0.19 & 2.36$\pm$0.12 & 37.82$\pm$1.05 & 1.16 & 0.49 & 40.74 & 46.72 & Q \\
19 & 247.2566 & 40.5349 & ETS + Bar & 16.71$\pm$0.83 & 10.16$\pm$0.51 & 47.62$\pm$0.54 & 2.34 & 0.72 & 47.69 & 49.16 & SF \\
20 & 247.0482 & 39.8219 & ETS + Bar & 12.24$\pm$0.61 & 7.76$\pm$0.39 & 14.80$\pm$0.44 & 0.70 & 0.28 & 174.36 & 40.29 & GV \\
21 & 247.1635 & 39.5185 & ETS + Bar & 7.51$\pm$0.34 & 4.58$\pm$0.20 & 55.32$\pm$1.06 & 1.44 & 0.49 & 50.71 & 37.54 & GV \\
22 & 246.9945 & 39.6040 & S0 + Bar & 5.20$\pm$0.26 & 3.16$\pm$0.16 & 7.49$\pm$1.12 & 1.16 & 0.38 & 5.90 & 24.94 & Q \\
23 & 247.0584 & 40.3138 & ETS + Bar & 7.84$\pm$0.28 & 4.77$\pm$0.17 & 0.75$\pm$0.55 & 0.86 & 0.38 & 4.23 & 32.49 & GV \\
24 & 247.3333 & 39.4290 & S0 + Bar & 10.85$\pm$0.48 & 6.60$\pm$0.29 & 164.87$\pm$0.91 & 1.55 & 0.46 & 165.63 & 33.92 & Q \\
25 & 246.3658 & 39.5520 & LTS + Bar & 6.23$\pm$0.31 & 4.06$\pm$0.20 & 161.55$\pm$1.03 & 0.74 & 0.37 & 15.36 & 34.92 & SF \\
26 & 247.3072 & 39.1818 & S0 + Bar & 6.10$\pm$0.31 & 3.72$\pm$0.19 & 51.65$\pm$0.90 & 1.95 & 0.60 & 47.66 & 44.33 & Q \\
27 & 246.9956 & 39.3029 & S0 + Bar & 9.15$\pm$0.23 & 5.57$\pm$0.14 & 155.47$\pm$0.85 & 1.51 & 0.55 & 159.97 & 27.25 & Q \\
28 & 246.3929 & 38.9333 & S0 + Bar & 5.01$\pm$0.13 & 4.02$\pm$0.10 & 117.52$\pm$1.14 & 0.96 & 0.27 & 174.38 & 45.73 & Q \\
29 & 247.4498 & 39.7162 & S0 + Bar & 3.16$\pm$0.05 & 1.92$\pm$0.05 & 6.58$\pm$0.71 & 0.82 & 0.27 & 11.06 & 21.21 & GV \\
30 & 246.9328 & 39.7438 & S0 + Bar & 3.82$\pm$0.10 & 2.34$\pm$0.06 & 74.39$\pm$0.99 & 2.09 & 0.72 & 83.20 & 33.04 & Q \\
31 & 247.8539 & 39.6930 & S0 + Bar & 4.02$\pm$0.10 & 2.45$\pm$0.06 & 38.61$\pm$1.09 & 0.98 & 0.32 & 32.06 & 38.44 & SF \\
32 & 247.1019 & 38.9331 & S0 + Bar & 1.82$\pm$0.05 & 1.10$\pm$0.03 & 128.96$\pm$1.36 & 0.78 & 0.31 & 130.37 & 34.12 & GV \\
33 & 247.6983 & 39.9784 & S0 + Bar & 4.44$\pm$0.22 & 2.70$\pm$0.13 & 33.06$\pm$0.89 & 1.30 & 0.46 & 35.99 & 31.58 & Q \\
34 & 247.2134 & 39.7555 & S0 + Bar & 3.49$\pm$0.15 & 2.16$\pm$0.10 & 1.28$\pm$0.92 & 0.86 & 0.34 & 166.80 & 38.45 & Q \\
35 & 247.3814 & 40.4128 & ETS + Bar & 7.24$\pm$0.29 & 4.44$\pm$0.18 & 93.10$\pm$0.87 & 1.01 & 0.36 & 101.63 & 40.74 & SF \\
36 & 246.8421 & 39.1096 & ETS + Bar & 4.43$\pm$0.22 & 2.91$\pm$0.15 & 112.76$\pm$1.46 & 0.50 & 0.22 & 151.27 & 33.32 & SF \\
37 & 247.0670 & 40.1159 & S0 + Bar & 7.49$\pm$0.26 & 4.58$\pm$0.16 & 177.84$\pm$1.11 & 0.86 & 0.33 & 8.44 & 29.09 & Q \\
38 & 247.5839 & 39.8043 & S0 + Bar & 6.36$\pm$0.32 & 3.91$\pm$0.19 & 103.52$\pm$0.74 & 1.74 & 0.54 & 114.24 & 36.66 & Q \\
39 & 247.1554 & 39.6989 & ETS + Bar & 10.14$\pm$0.28 & 6.17$\pm$0.17 & 130.95$\pm$0.76 & 1.31 & 0.57 & 128.01 & 42.97 & GV \\ \bottomrule \bottomrule
\end{tabular}
}\tablefoot{[1] Bar ID is the serial number of the galaxies in the sample; [2] right ascension (RA) in degrees; [3] declination (Dec) in degrees; [4] visual morphology; [5] projected bar radius ($R_{\mathrm{bar}}^{\mathrm{proj}}$) in arcsec; [6] deprojected bar radius ($R_{\mathrm{bar}}^{\mathrm{deproj}}$) in kpc; [7] bar position angle (east of north) in degrees; [8] normalised bar radius with R$_{\mathrm{50}}$; [9] normalised bar radius with R$_{\mathrm{90}}$; [10] galaxy position angle (east of north) in degrees; [11] inclination angle of galaxy in degrees; [12] star-formation rate class (SFR class; \citep{2014SerAJ.189....1S}): star-forming (SF; \(\log(\mathrm{sSFR}) \geq -10.8\)), green valley (GV; \(-10.8 < \log(\mathrm{sSFR}) < -11.8\)), and quenched (Q; \(\log(\mathrm{sSFR}) \leq -11.8\)).}
\end{table*}

\subsubsection{Structural Decomposition} 
We performed two-dimensional photometric decompositions of the barred galaxies using \textsc{GALFIT} (version 3.0; \citealt{2002AJ....124..266P,2010AJ....139.2097P}). Each galaxy was modelled as a three-component Sérsic configuration comprising a bulge, disc, and bar. This approach explicitly separates the bar contribution from the bulge component, thereby reducing the risk that bar light artificially inflates the bulge-to-total luminosity ratio ($B/T$) or the bulge Sérsic index ($n_{\mathrm{bulge}}$). The derived structural parameters, including $B/T$, $n_{\mathrm{bulge}}$, the bar's effective radius, and the bar-to-total ratio, provide complementary diagnostics of bar structure within the cluster environment.

To validate our procedure, we cross-matched our barred sample with the five-band SDSS decomposition catalogue of \citet{2018MNRAS.473.4731K}, identifying 14 galaxies in common. For these galaxies, we performed independent $r$-band decompositions and verified that the resulting structural parameters are consistent with those reported in their study. The same methodology was then applied to the remaining 25 barred galaxies. An example of the three-component decomposition (bulge, disc, and bar) for Bar ID 4 is shown in Fig.~\ref{fig:galfit}. As an additional consistency check, we compared the radial surface-brightness profiles of the original images and best-fitting models, finding good agreement in all cases. The resulting structural parameters are listed in Table~\ref{tab:bar_decomp}. 

To investigate environmental trends, we examine the bar radius normalised by three independent measures of galaxy size: the disc scale length ($h_{\mathrm{disc}}$), and the radii enclosing 50\% and 90\% of the total flux ($R_{50}$ and $R_{90}$), adopted from the NSA catalogue (Fig.~\ref{fig:bar_var_disc} and  \ref{fig:bar_parameters}). The disc scale length is derived from the disc effective radius, using $R_{\mathrm{e,disc}}$ = 1.678$h_{\mathrm{disc}}$ \citep{2005PASA...22..118G, 2016MNRAS.458.1199E}. We analyse these quantities as a function of projected cluster-centric distance for all morphologies combined, as well as separately for ETS+Bar and S0+Bar galaxies (Fig.~\ref{fig:bar_var_disc}). For ETS+Bar galaxies, the cluster-centric behaviour is consistent across all three normalisations: the normalised bar radius appears to be enhanced towards the cluster centre, decreases at intermediate radii, and rises again in the outer regions. The consistency of this behaviour across different size scalings suggests that the trend is not driven by the specific choice of normalisation.

In contrast, S0+Bar galaxies do not show a clear variation of normalised bar size with cluster-centric distance. This difference suggests that the environmental dependence of bar structure may be morphology-dependent, with ETS+Bar galaxies potentially exhibiting a stronger response to cluster environment.

To assess the significance of the observed non-monotonic behaviour, we performed a bootstrap-based analysis by fitting a quadratic relation to the unbinned data. We find that the curvature is more pronounced for ETS+Bar galaxies, while the S0+Bar sample does not show a statistically robust trend. However, given the limited sample size and substantial scatter, these results should be considered suggestive.

\begin{figure*}[htb!]
     \centering
     \begin{subfigure}[b]{0.45\textwidth}
         \centering
         \includegraphics[width=\textwidth]{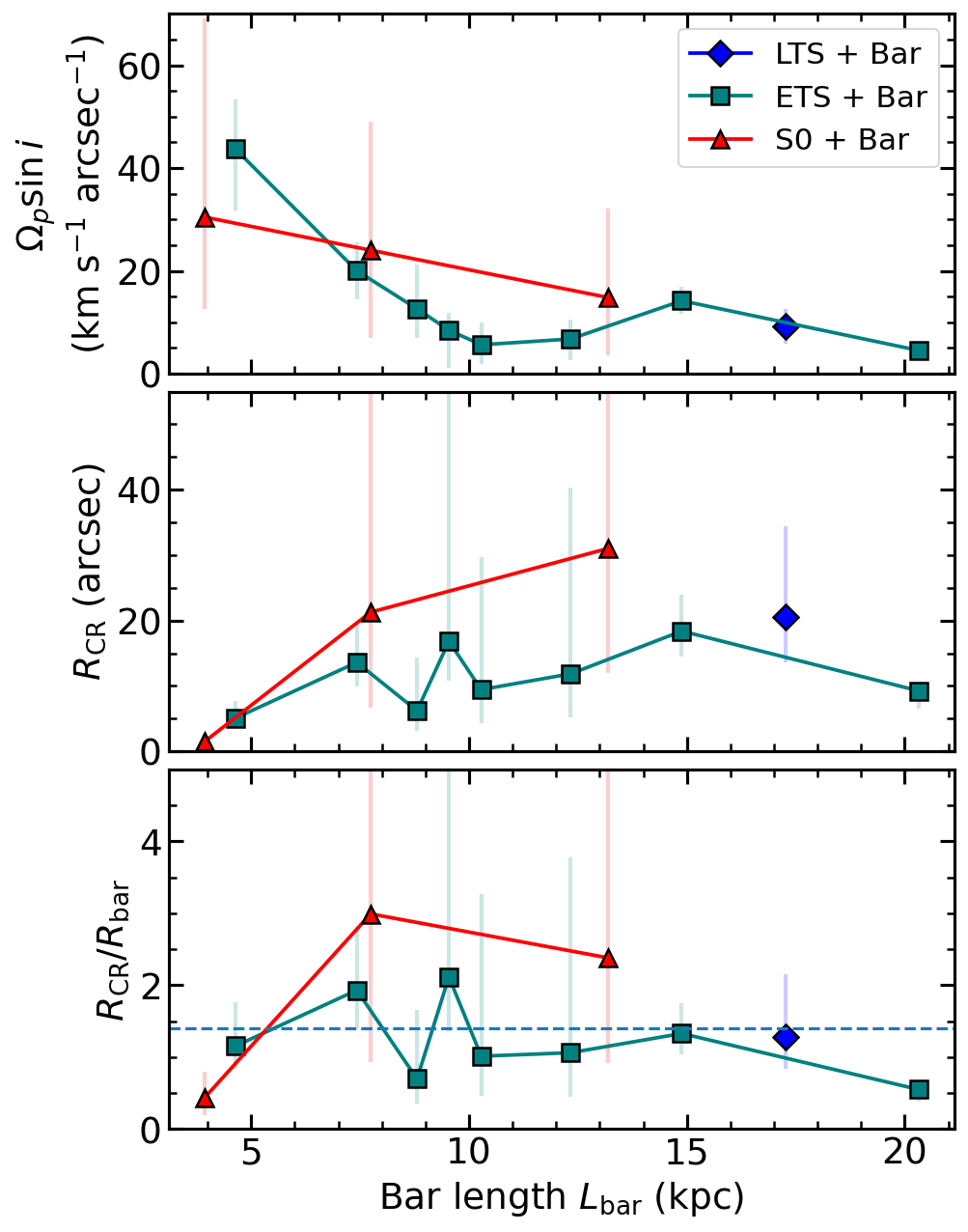}
     \end{subfigure}
     \begin{subfigure}[b]{0.45\textwidth}
         \centering
         \includegraphics[width=\textwidth]{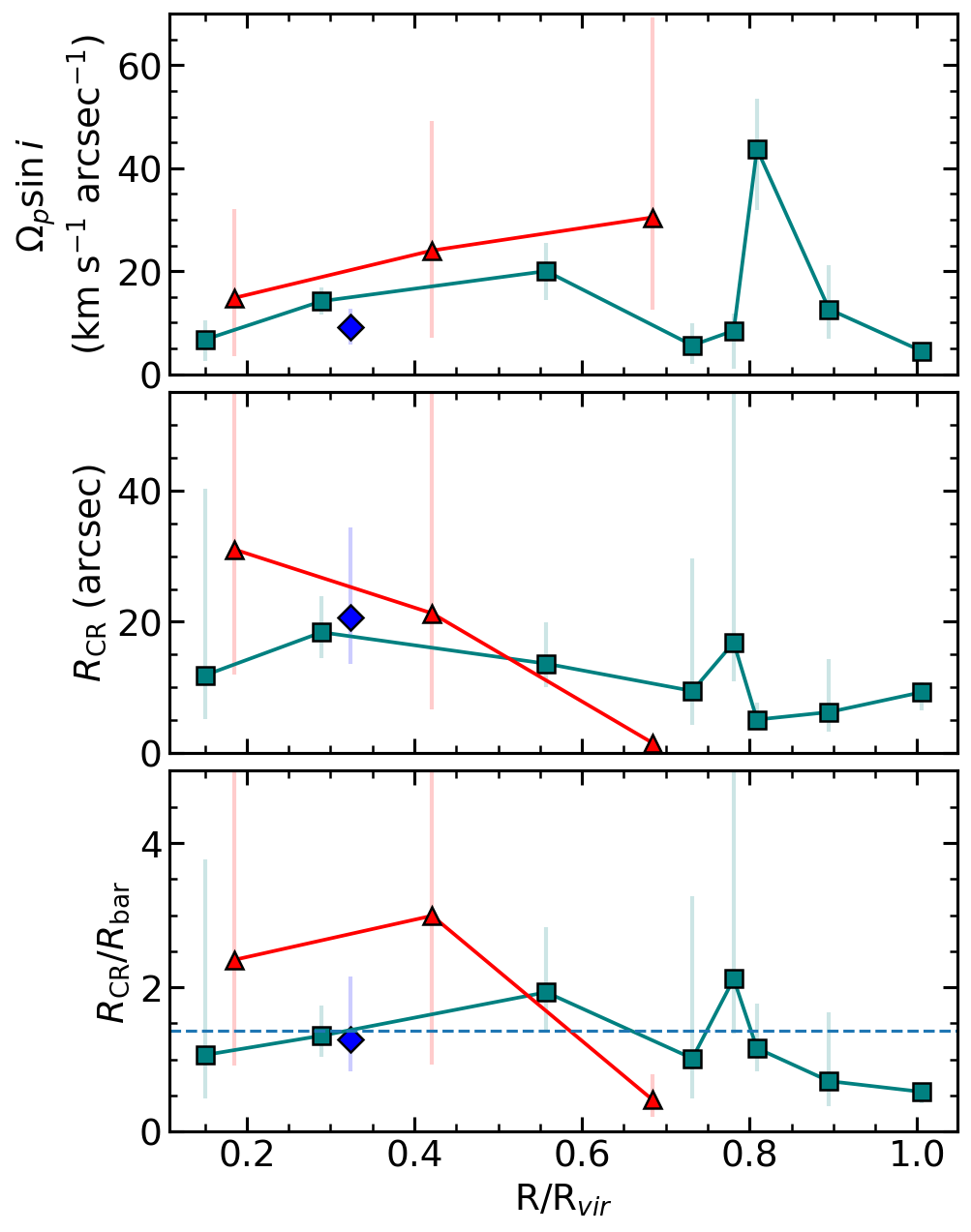}
     \end{subfigure}
     \caption{Bar dynamical properties as a function of bar size and projected cluster-centric distance. (top panels) bar pattern speed ($\Omega_{\mathrm{p}}$). (middle panels) corotation radius ($R_{\mathrm{cr}}$). (bottom panels) dimensionless ratio $\mathcal{R} = R_{\mathrm{cr}}/R_{\mathrm{bar}}$. The left column shows these quantities as a function of projected bar radius, while the right column shows their variation with projected cluster-centric distance. The dashed horizontal line in the bottom panels marks $\mathcal{R} = 1.4$, commonly adopted as the boundary between fast and slow rotator bars. Symbols are color-coded by morphology: S0+Bar (red), ETS+Bar (teal), and LTS+Bar (blue).}

    \label{fig:bar_parameters_p}
\end{figure*}

In Sect.~\ref{sec:sf}, we identified a transition in the fractions of star-forming and quenched galaxies at a projected cluster-centric distance of approximately 1\,Mpc. A comparable change in the behaviour of the normalised bar radius is also observed at a similar scale. 

The precise location of this transition may depend on radial binning, we repeated the analysis with bin widths ranging from 0.2 to 0.8\,Mpc. Across all binning choices, the intersection of the star-forming and quenched fractions consistently lies at $1.03 \pm 0.06$\,Mpc, corresponding closely to $\sim$0.5\,$R_{\mathrm{vir}}$ ($\sim$1.14\,Mpc). Motivated by this robustness, we adopt 1.14\,Mpc as a characteristic scale and divide the cluster into inner and outer regions for subsequent analysis. Applying this criterion to the barred sample yields 20 galaxies in the inner region and 19 in the outer region.

\subsubsection{Bar Pattern Speed}
Of the 39 barred galaxies in our sample, 22 have available MaNGA integral-field spectroscopy. For these galaxies, structural parameters and deprojected bar lengths have already been derived. We now investigate the dynamical state of these bars by measuring their pattern speeds.

Bar pattern speeds are estimated using the publicly available implementation of the Tremaine--Weinberg (TW) method \citep{1984ApJ...282L...5T} developed by \citet{geron_galaxy_2023}, who applied the method to 225 MaNGA galaxies. Owing to differences in sample selection, only four galaxies overlap between their catalogue and ours.

Before applying the TW method, we verified that the necessary geometric conditions are satisfied. The method requires inclinations in the range $20^{\circ} < i < 70^{\circ}$ and that the bar position angle differs by more than $10^{\circ}$ from both the photometric major and minor axes. The kinematic position angles were determined from MaNGA stellar velocity fields using the \textsc{PaFit} v2.0.8 package \citep{2006MNRAS.366..787K}. All 22 galaxies satisfy the inclination requirement. Two galaxies fail the position-angle criterion and are excluded, leaving 20 galaxies.

We applied the TW method to these 20 galaxies and adopted the same quality threshold as \citet{geron_galaxy_2023}, selecting only measurements with a normalised root-mean-square error (NRMSE) $< 0.2$. This yields a final subsample of 12 galaxies with reliable pattern-speed estimates, with uncertainties estimated using Monte Carlo resampling and are shown in Fig.~\ref{fig:bar_parameters_p}

\begin{figure*}[htb!]
     \centering
     \begin{subfigure}[b]{0.48\textwidth}
         \centering
         \includegraphics[width=\textwidth]{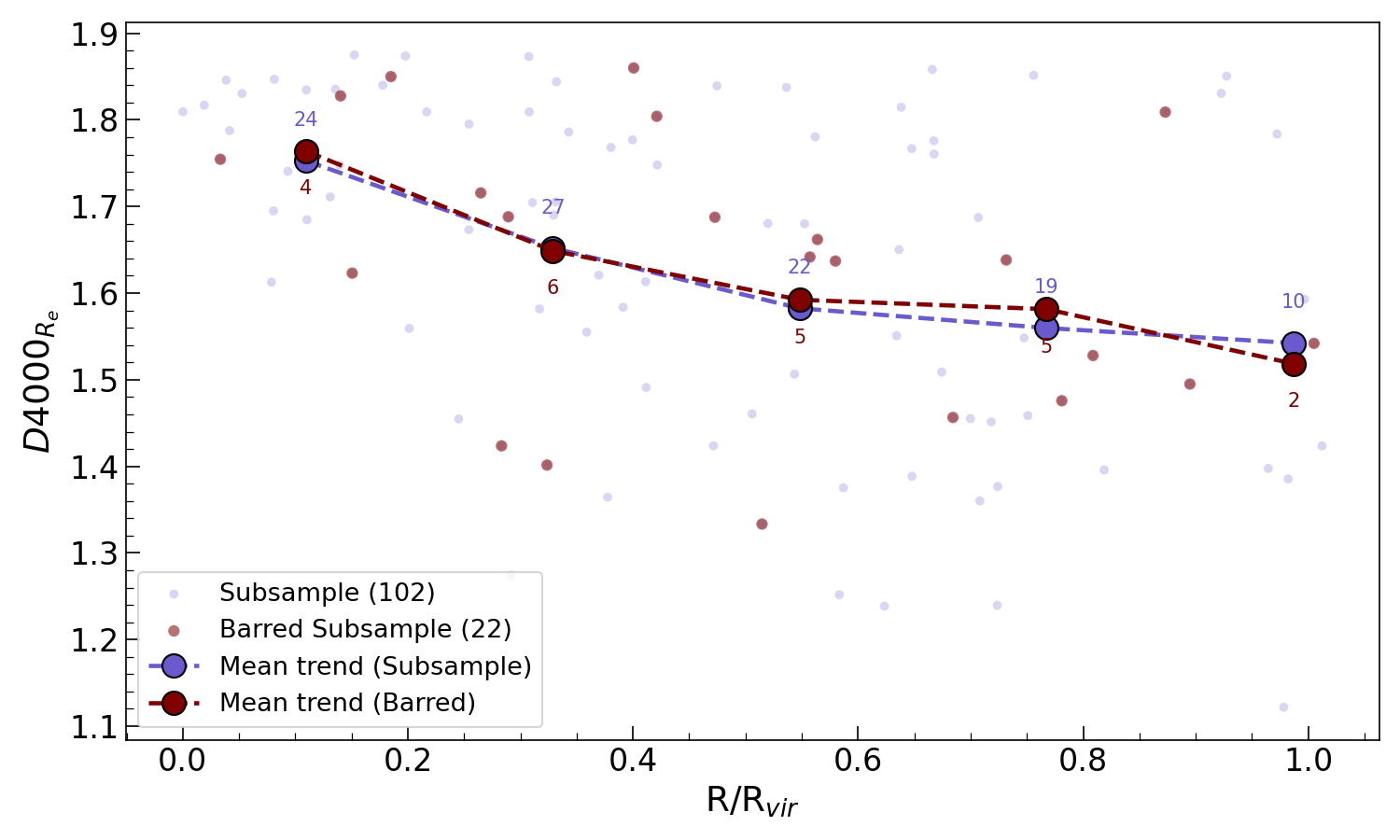}
     \end{subfigure}
     \begin{subfigure}[b]{0.48\textwidth}
         \centering
         \includegraphics[width=\textwidth]{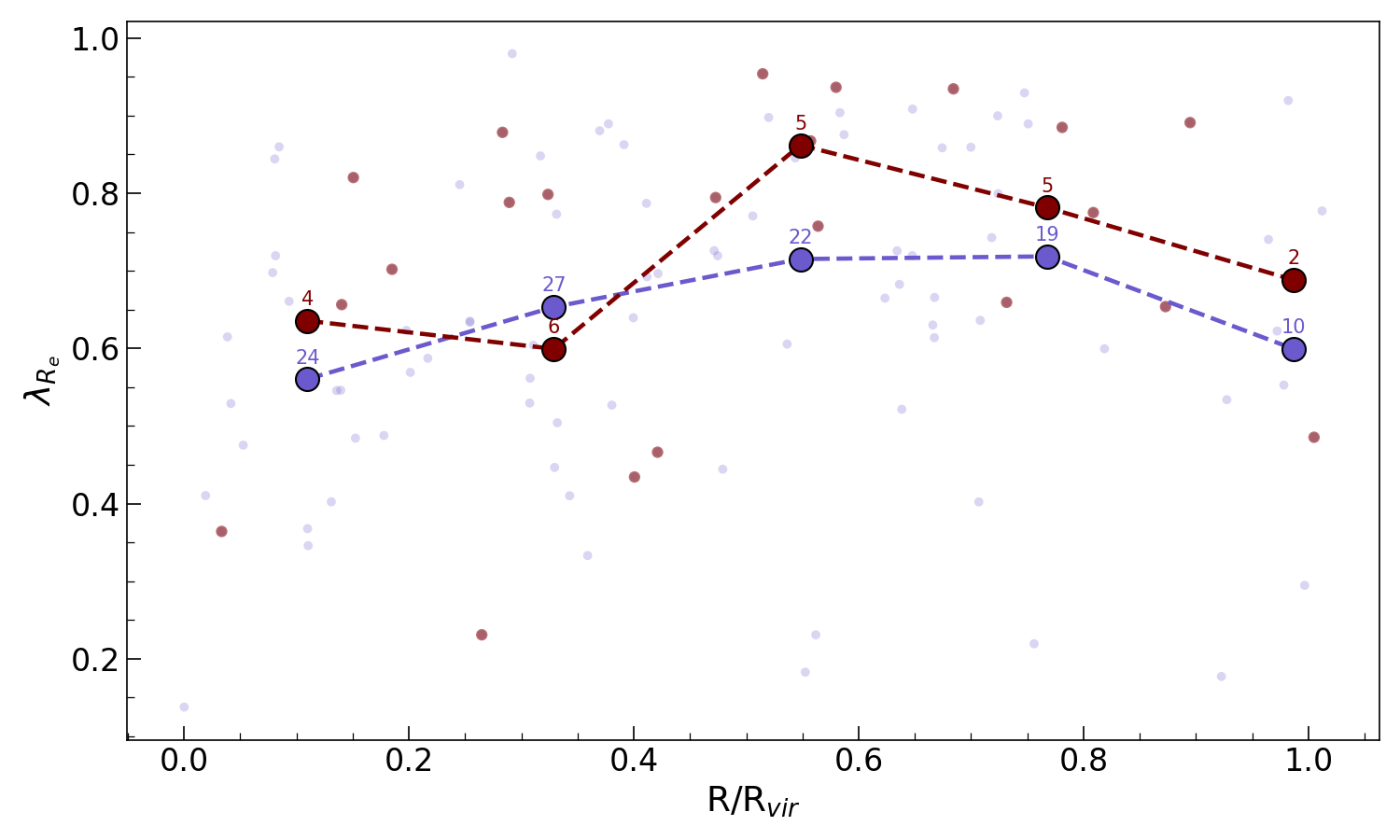}
     \end{subfigure}
     \caption{Cluster-centric trends of stellar population age and projected specific angular momentum. (left) $D4000_{R_{\mathrm{e}}}$ as a function of projected cluster-centric distance. (right) $\lambda_{R_{\mathrm{e}}}$ as a function of projected cluster-centric distance. The full MaNGA subsample (102 galaxies) is shown by the solid purple line, while the barred-galaxy subsample (22 galaxies) is shown by the dashed maroon line. Points represent bin-averaged values computed using a radial bin size of 0.5\,Mpc, and the numbers adjacent to each bin indicate the number of galaxies included.}
        \label{fig:ss_var}
\end{figure*}

For these galaxies, we derive the bar pattern speed ($\Omega_{\mathrm{p}}$), the corotation radius ($R_{\mathrm{cr}}$), and the dimensionless ratio $\mathcal{R} = R_{\mathrm{cr}}/R_{\mathrm{bar}}$ (Table~\ref{tab:bar_decomp}; Fig.~\ref{fig:bar_parameters_p}). These quantities enable us to examine the dynamical state of bars and their variation with cluster-centric distance.

We find that the cluster-centric behaviour of $\Omega_{\mathrm{p}}$ for ETS+Bar galaxies, broadly follows the trends observed in normalised bar radius. In ETS+Bar galaxies, bars located closer to the cluster center appear to exhibit larger normalized radii and lower pattern speeds, while galaxies at intermediate radii show smaller normalized bar sizes and comparatively higher pattern speeds. Most galaxies lie close to the canonical boundary between fast and slow rotators ($\mathcal{R} \approx 1.4$), although with substantial scatter owing to the limited sample size. 

Motivated by the transition scale identified in Sect.~\ref{sec:sf}, we examine the barred galaxies separately in inner and outer regions defined at 1.14\,Mpc ($\sim 0.5\,R_{\mathrm{vir}}$). The structural decomposition results (Table~\ref{tab:inner_outer_all}) show that the median projected bar effective radius and bar radius is smaller for ETS+Bar galaxies in the inner region, whereas the normalised bar radius remains larger when scaled by $h_{\mathrm{disc}}$, $R_{50}$, or $R_{90}$. This suggests that the environmental signal primarily reflects changes in bar size relative to the host galaxy's scale rather than absolute bar size alone. In contrast, S0+Bar galaxies show weaker or absent trends in normalised bar size, although their pattern speeds display a mild increase with cluster-centric distance. This trend is not significant, as it is based on only three data points.

The bulge structural parameters provide additional context. The Sérsic indices are generally consistent with pseudo-bulge-like values, suggesting an important role for secular evolution. Both the bar-to-total and bulge-to-total luminosity ratios are, on average, higher in the inner region than in the outer region (Fig.~\ref{fig:box_plot}), although bulges remain the dominant component overall. When separated by morphology, bars are relatively more prominent in ETS+Bar galaxies than in S0+Bar galaxies in the inner region, while the opposite tendency is observed in the outskirts.

To assess whether this transition is physically motivated, we tested a range of radial cuts and find that the increase in bar size remains stable over $\sim 0.96$--$1.32$\,Mpc ($\sim 0.42$--$0.58\,R_{\mathrm{vir}}$). This indicates that the result is not sensitive to the exact choice of threshold and reflects a physically motivated transition scale.

Taken together, these results suggest a morphology-dependent environmental influence on bar structure and dynamics. The trends are appear to be more clearly defined in ETS+Bar galaxies, whereas S0+Bar galaxies exhibit weaker variations. While the limited sample size precludes definitive conclusions, the observed behaviour is qualitatively consistent with recent studies of barred galaxies in dense environments \citep{2026A&A...705A.115C,2026MNRAS.547ag175P} and with simulations indicating that tidal interactions can modify bar properties in cluster environments \citep{2025A&A...702A...7L}. 

We note that some outer-region galaxies display comparatively large normalised bar radii and low pattern speeds. These may represent galaxies that formed strong bars prior to cluster infall, although alternative explanations, such as differences in gas content or orbital history, cannot be excluded.

While we observe correlations between bar properties and cluster-centric distance, these do not directly establish causality. We find no strong dependence on stellar mass, as galaxies span a broad mass range with significant scatter, consistent with the dynamically evolving state of the cluster. The observed trends may reflect the influence of environmental processes acting during infall, such as tidal interactions, which can affect bar evolution, particularly in ETS+Bar galaxies. The different behaviour observed in the outer regions may be associated with galaxies that are recently accreted, although alternative scenarios, such as pre-processing in group environments prior to cluster infall \citep{2023A&A...678A.147L}, may also contribute. These results should therefore be interpreted as indicative of possible environmental influence rather than definitive evidence of a specific mechanism.

Another factor that may play a role on bar structures in cluster galaxies is the tidal field of the cluster, which varies strongly with radius towards the center of the cluster \citep{valluri.1993}. This may cause kinematic heating of the stellar discs, which in turn will affect bar formation and structure.

\subsection{Angular Momentum and Stellar Populations}
We next examine the subsample of 102 MaNGA galaxies for which measurements of $D4000_{R_{\mathrm{e}}}$ and $\lambda_{R_{\mathrm{e}}}$ are available from the Pipe3D catalogue. The $D4000$ index measures the strength of the 4000\,\AA\ break and is widely used as an age-sensitive indicator of stellar populations \citep{1999ApJ...527...54B}. Following \citet{2003MNRAS.346.1055K} and \citet{2018ApJ...867..118K}, we adopt $D4000_{R_{\mathrm{e}}} = 1.5$ as a practical division between younger and older stellar populations. The parameter $\lambda_{R_{\mathrm{e}}}$ serves as a proxy for the projected stellar specific angular momentum within one effective radius, providing a link between galaxy kinematics and structural evolution \citep{2018MNRAS.477.4711G}.

We first examine the dependence of these quantities on projected cluster-centric distance (Figure~\ref{fig:ss_var}). Galaxies located closer to the cluster centre tend to exhibit higher $D4000_{R_{\mathrm{e}}}$ values and lower $\lambda_{R_{\mathrm{e}}}$, while galaxies in the outskirts show comparatively younger stellar populations and higher projected angular momentum. Although the scatter is substantial, the overall behaviour is consistent with an environmental influence on quenching and dynamical evolution towards the cluster core.

Figure~\ref{fig:ss_var} also shows the corresponding trends for the barred subsample. Within the uncertainties, barred galaxies exhibit broadly similar cluster-centric behaviour to the full MaNGA sample, without evidence for a systematically distinct global trend.

To isolate environmental effects more clearly, we divide the barred galaxies into inner and outer regions using the 1.14\,Mpc threshold defined in Sect. ~\ref{sec:bar}. In this framework (Table~\ref{tab:inner_outer_all}), barred galaxies in the inner region tend to show lower $\lambda_{R_{\mathrm{e}}}$ and higher $D4000_{R_{\mathrm{e}}}$ than those in the outer region. When separated by morphology, ETS+Bar galaxies in the inner region display higher angular momentum and younger stellar populations than S0+Bar galaxies at comparable radii. In the outer region, ETS+Bar and S0+Bar exhibit similar $\lambda_{R_{\mathrm{e}}}$ values, although ETS+Bar galaxies remain, on average, younger than their S0+Bar counterparts.

Recent MaNGA-based work by \citet{2026MNRAS.546ag015A} reports that barred galaxies preferentially exhibit lower stellar angular momentum, consistent with angular-momentum redistribution driven by secular evolution. Our results are qualitatively consistent with this picture and suggest that the cluster environment modulates these properties. In particular, barred galaxies in the inner cluster region tend to be dynamically cooler and host older stellar populations than those in the outskirts. The persistence of younger stellar populations in ETS+Bar galaxies relative to S0+Bar at similar radii may indicate differences in evolutionary stage, with S0+Bar galaxies representing more advanced quenching within the cluster environment.

Given the limited number of barred MaNGA galaxies, these trends should be regarded as indicative rather than definitive; nevertheless, they reinforce the morphology-dependent environmental signatures identified in the bar-structure analysis.

\section{Conclusions and Summary}
We investigate the environmental imprint on galaxy evolution in the dynamically unrelaxed cluster Abell~2199 using a hierarchical data set. Our analysis is based on a parent sample of 578 spectroscopically confirmed cluster members and a master sample of 325 galaxies with homogeneous stellar mass and SFR measurements, of which 314 also have local environmental density estimates. From this master sample, we identify 39 barred galaxies and a MaNGA subsample of 102 galaxies with integral-field spectroscopy. Of these, 22 barred galaxies have MaNGA coverage, and 12 yield reliable bar pattern-speed measurements using the TW method. Although the barred-galaxy subsample is necessarily modest in size, the observed trends with cluster-centric distance are broadly consistent within the uncertainties.

Our primary goal is to quantify the interplay between bar structure, bar dynamics, and the cluster environment. Bar structural properties are derived from ellipse fitting and three-component (bulge+disc+bar) photometric decompositions. The global cluster properties are first characterised through star formation activity and morphology, providing the environmental framework for the bar analysis. Our main findings are summarised as follows:

\begin{enumerate}
    \item The cluster exhibits a star formation--density relation: the central regions are dominated by quenched galaxies, whereas the outskirts are preferentially populated by star-forming galaxies, with a noticeable concentration of green valley galaxies towards the cluster centre.
    
    \item Despite its non-relaxed dynamical state, Abell~2199 shows a morphology--density relation. The fraction of S0 galaxies increases towards the cluster centre, while ETS become more prevalent at larger cluster-centric distances.
    
    \item For barred ETS galaxies, the normalised bar radius (scaled by $h_{\mathrm{disc}}$, $R_{50}$, or $R_{90}$) shows a cluster-centric variation: bars are relatively more extended towards the cluster centre, decrease in relative size at intermediate radii, and increase again in the outskirts. This behaviour is not clearly present in S0+Bar galaxies.
    
    \item The bar pattern speed displays a broadly complementary behaviour for ETS galaxies, with lower values in the central region and higher values at intermediate radii. Most galaxies lie close to the canonical boundary between fast and slow rotator bars, albeit with substantial scatter due to the limited sample size.
    
    \item Motivated by a transition near $\sim$0.5\,$R_{\mathrm{vir}}$, we divide the cluster into inner and outer regions. This separation appears to enhance the environmental signal: in the inner region, bars are relatively more prominent in ETS than in S0 galaxies, whereas the contrast weakens or reverses in the outer region.
    
    \item Stellar populations become progressively older, and projected angular momentum decreases towards the cluster centre. Within the barred subsample, ETS galaxies retain higher angular momentum and younger stellar populations than S0 galaxies at comparable radii, indicating a morphology-dependent response to the cluster environment.
\end{enumerate}

Taken together, these results suggest that the cluster environment may influence both the global evolutionary state of galaxies and the internal secular processes traced by bars. In particular, ETS galaxies appear to exhibit a stronger environmental coupling between bar structure, bar dynamics, and stellar populations than S0 galaxies. Although based on a modest sample of barred galaxies with dynamical measurements, the observed trends indicate that bar properties provide a sensitive probe of environmental influence across the cluster-centric distance.

Overall, this study suggests an important role of the cluster environment in shaping the structural and dynamical properties of barred galaxies. However, the trends reported here may also reflect the specific dynamical state of Abell 2199, a non-relaxed cluster. Future work will extend this analysis to clusters with different dynamical states and explore these trends using cosmological simulations.

\begin{acknowledgements}
 We thank the anonymous referee for the thoughtful review, which has improved the clarity of this work. MD and SB acknowledge the support of the Science and Engineering Research Board (SERB) Core Research Grant CRG/2022/004531 and the Department of Science and Technology (DST) grant DST/WIDUSHIA/PM/2023/25(G) for this research. EL{\L} was supported by the National Science Centre of Poland with grant 2025/57/B/ST9/00321. Funding for the Sloan Digital Sky Survey IV has been provided by the Alfred P. Sloan Foundation, the U.S. Department of Energy Office of Science, and the Participating Institutions. SDSS-IV acknowledges support and resources from the Center for High-Performance Computing at the University of Utah. The SDSS web site is www.sdss.org. SDSS-IV is managed by the Astrophysical Research Consortium for the Participating Institutions of the SDSS Collaboration including the Brazilian Participation Group, the Carnegie Institution for Science, Carnegie Mellon University, the Chilean Participation Group, the French Participation Group, Harvard-Smithsonian Center for Astrophysics, Instituto de Astrof\'isica de Canarias, The Johns Hopkins University, Kavli Institute for the Physics and Mathematics of the Universe (IPMU) / University of Tokyo, Lawrence Berkeley National Laboratory, Leibniz Institut f\"ur Astrophysik Potsdam (AIP), Max-Planck-Institut f\"ur Astronomie (MPIA Heidelberg), Max-Planck-Institut f\"ur Astrophysik (MPA Garching), Max-Planck-Institut f\"ur Extraterrestrische Physik (MPE), National Astronomical Observatories of China, New Mexico State University, New York University, University of Notre Dame, Observat\'ario Nacional / MCTI, The Ohio State University, Pennsylvania State University, Shanghai Astronomical Observatory, United Kingdom Participation Group, Universidad Nacional Aut\'onoma de M\'exico, University of Arizona, University of Colorado Boulder, University of Oxford, University of Portsmouth, University of Utah, University of Virginia, University of Washington, University of Wisconsin, Vanderbilt University, and Yale University. The Legacy Surveys consist of three individual and complementary projects: the Dark Energy Camera Legacy Survey (DECaLS; Proposal ID 2014B-0404; PIs: David Schlegel and Arjun Dey), the Beijing-Arizona Sky Survey (BASS; NOAO Prop. ID 2015A-0801; PIs: Zhou Xu and Xiaohui Fan), and the Mayall z-band Legacy Survey (MzLS; Prop. ID 2016A-0453; PI: Arjun Dey). DECaLS, BASS and MzLS together include data obtained, respectively, at the Blanco telescope, Cerro Tololo Inter-American Observatory, NSF’s NOIRLab; the Bok telescope, Steward Observatory, University of Arizona; and the Mayall telescope, Kitt Peak National Observatory, NOIRLab. Pipeline processing and analyses of the data were supported by NOIRLab and the Lawrence Berkeley National Laboratory (LBNL). The Legacy Surveys project is honored to be permitted to conduct astronomical research on Iolkam Du’ag (Kitt Peak), a mountain with particular significance to the Tohono O’odham Nation. This research has made use of NASA's Astrophysics Data System (ADS), SIMBAD database and the NASA/IPAC Extragalactic Database (NED).     
\end{acknowledgements}

\bibliographystyle{bibtex/aa} 
\bibliography{bibtex/sample} 

\clearpage

\begin{appendix}
\onecolumn
\section{Additional Figures and Tables}

\begin{figure*}[ht!]
    \sidecaption
    \includegraphics[width=0.31\textwidth]{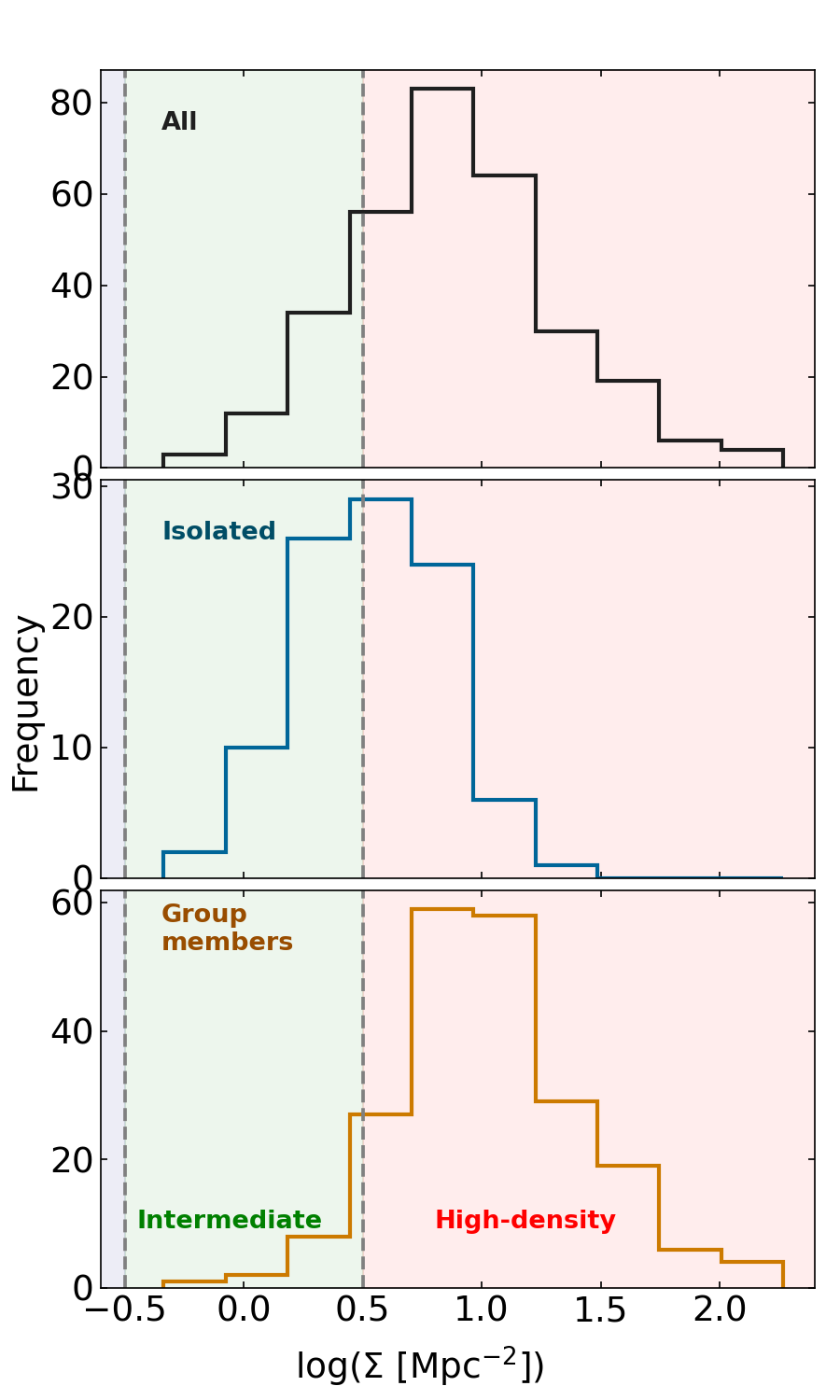}
    \hspace{0.01\textwidth}
    \includegraphics[width=0.31\textwidth]{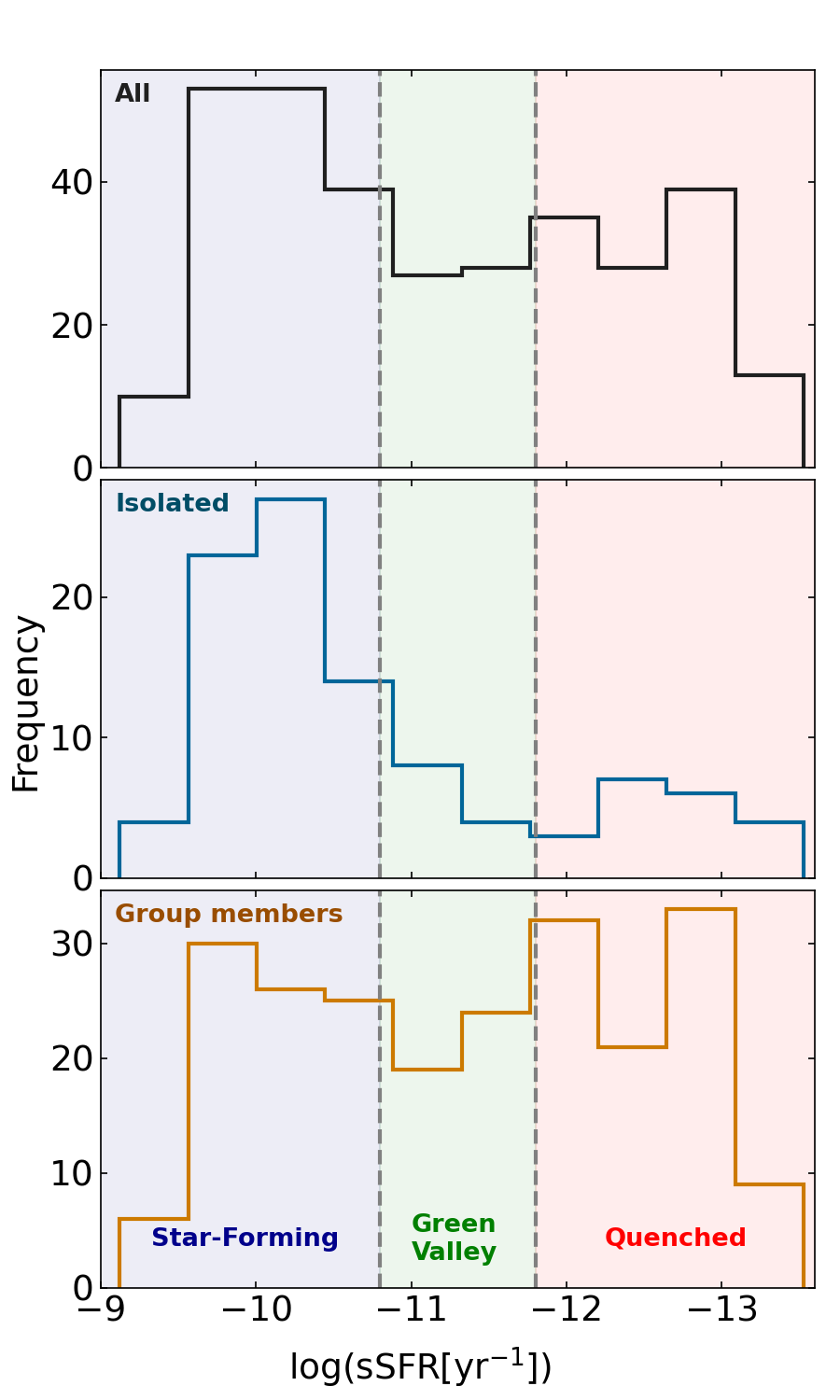}
     \caption{Distribution of local environmental density and specific star formation rate (sSFR) for the master sample. (left) Distribution of local surface density, $\log \Sigma$ (Mpc$^{-2}$), for the full master sample (top), isolated galaxies (middle), and group members (bottom). Shaded regions indicate the adopted intermediate- and high-density regimes. (right) Distribution of $\log(\mathrm{sSFR}\,(\mathrm{yr}^{-1}))$ for the same three subsets. Shaded regions denote the star-forming, green valley, and quenched classifications. Isolated galaxies are defined as systems not associated with any identified group, while group members correspond to galaxies assigned to groups within the cluster according to \citet{2017ApJ...842...88S}.}
    \label{fig:sample}
\end{figure*}

\begin{figure}[ht!]
    \sidecaption
    \includegraphics[width=0.5\columnwidth]{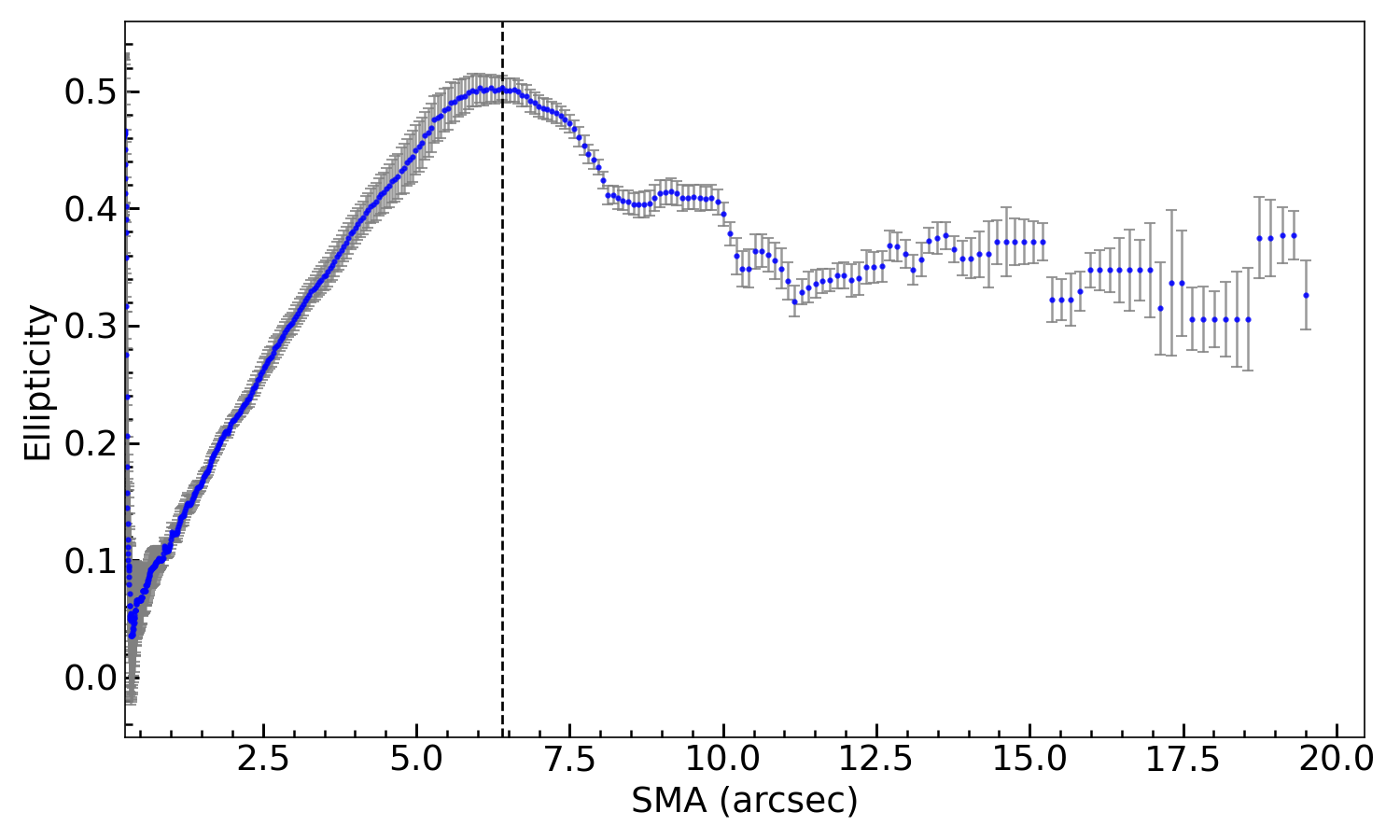}
    \caption{Isophotal ellipticity profile as a function of semi-major axis (SMA) for galaxy Bar ID 4, derived from elliptical isophote fitting of the SDSS $r$-band image. The dashed vertical line marks the projected bar radius, defined as the semi-major axis at which the ellipticity reaches its maximum.}
    \label{fig:ellip}
\end{figure}

\begin{figure*}[ht!]
    \centering
    \includegraphics[width=0.8\textwidth]{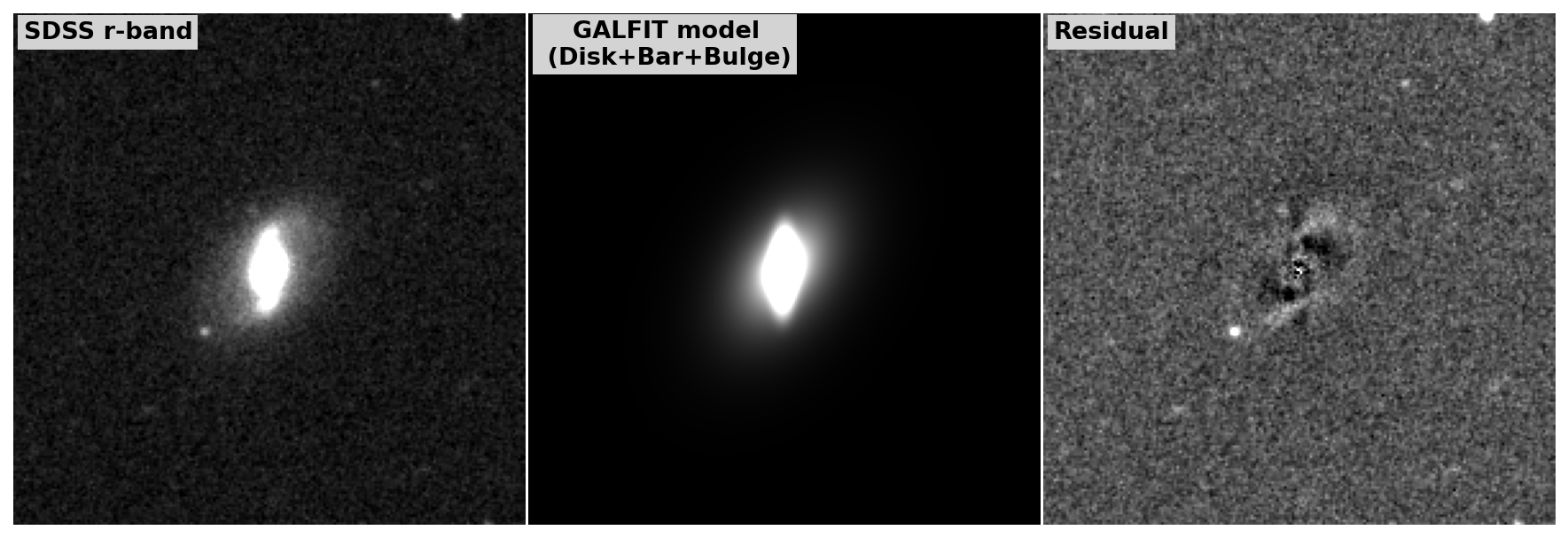}
    \caption{(left) SDSS $r-$ band image, (middle) GALFIT model, and corresponding (right) residual for Bar ID 4.}
    \label{fig:galfit}
\end{figure*}

\begin{table*}[ht!]
\centering
\caption{2D photometric decomposition parameters and dynamical properties for the barred galaxy sample.}
\label{tab:bar_decomp}
\resizebox{\columnwidth}{!}{%
\begin{tabular}{@{}lllllllllllll@{}}
\toprule \toprule
Bar ID &
\makecell{$\mathrm{mag}_{\mathrm{disc}}$} &
\makecell{$\mathrm{R}_{\mathrm{e,disc}}$ \\ (kpc)} &
\makecell{$\mathrm{mag}_{\mathrm{bar}}$} &
\makecell{$\mathrm{R}_{\mathrm{e,bar}}$ \\ (kpc)} &
\makecell{$\mathrm{n}_{\mathrm{bar}}$} &
\makecell{$\mathrm{mag}_{\mathrm{bulge}}$} &
\makecell{$\mathrm{R}_{\mathrm{e,bulge}}$ \\ (kpc)} &
\makecell{$\mathrm{n}_{\mathrm{bulge}}$} &
\makecell{$R_{\mathrm{bar}}/h_{\mathrm{disc}}$} &
\makecell{PlateID} &
\makecell{$\mathrm{PA}_{\mathrm{kin}}$ \\ (deg)} &
\makecell{$\Omega_{\mathrm{bar}}$ \\ (km\,s$^{-1}$\,arcsec$^{-1}$)}
\\
\midrule
1  & 16.15 & 5.68  & 17.53 & 2.65 & 0.23 & 18.02 & 0.51 & 0.85 & 0.94 &$-$         & $-$             & $-$                        \\
2  & 15.87 & 7.96  & 17.8  & 4.82 & 0.11 & 17.33 & 0.74 & 3.15 &  1.25 &$-$         & $-$             & $-$                        \\
3  & 15.25 & 5.95  & 16.74 & 1.77 & 0.56 & 17.15 & 0.66 & 0.88 &  0.37 &$-$         & $-$             & $-$                        \\
4  & 16.12 & 5.18  & 17.64 & 3.05 & 0.22 & 17.61 & 0.54 & 0.97 &  1.26 &$-$         & $-$             & $-$                        \\
5  & 13.49 & 23.34 & 16.68 & 4.93 & 0.44 & 16.37 & 0.80 & 1.72 &  0.62 &8602-12705  & $145.05\pm2.34$ & $9.15^{+3.50}_{-3.30}$       \\
6  & 16.36 & 3.78  & 18.75 & 1.41 & 0.45 & 18.18 & 0.35 & 1    &  1.05 &$-$         & $-$             & $-$                        \\
7  & 16.2  & 4.32  & 18.11 & 1.68 & 0.25 & 18.09 & 0.74 & 1.5  &  0.62 &$-$         & $-$             & $-$                        \\
8  & 16.4  & 3.59  & 17.34 & 2.77 & 0.15 & 16.77 & 0.47 & 1.05 &  1.61 &$-$         & $-$             & $-$                        \\
9  & 17.14 & 7.36  & 17.99 & 2.28 & 0.37 & 18.41 & 1.57 & 3.19 &  0.45 &9869-6102   & $55.86\pm14.86$ & $30.45^{+38.73}_{-17.88}$    \\
10 & 14.57 & 10.51 & 16.56 & 3.12 & 0.57 & 16.48 & 0.60 & 1.38 &  0.59 &8604-12703  & $98.92\pm1.62$  & $20.02^{+5.45}_{-5.54}$      \\
11 & 16.68 & 2.89  & 17.52 & 2.82 & 0.39 & 18.22 & 0.82 & 3.23 &  1.01 &$-$         & $-$             & $-$                        \\
12 & 14.91 & 8.48  & 16.65 & 3.42 & 0.5  & 16.23 & 0.54 & 1.1  &  1.02 &8603-12702  & $137.84\pm8.47$ & $5.61^{+4.37}_{-3.69}$       \\
13 & 15.96 & 2.76  & 18.28 & 1.27 & 0.67 & 17.19 & 0.30 & 0.79 &  1.37 &$-$         & $-$             & $-$                        \\
14 & 16.3  & 4.42  & 17.41 & 2.35 & 0.46 & 16.95 & 0.54 & 1.15 &  1.25 &$-$         & $-$             & $-$                        \\
15 & 17.02 & 3.42  & 18.25 & 1.66 & 0.63 & 18.41 & 0.35 & 0.91 &  1.22 &$-$         & $-$             & $-$                        \\
16 & 17.1  & 4.53  & 18.52 & 2.42 & 0.13 & 18.98 & 1.47 & 2.97 &  1.05 &$-$         & $-$             & $-$                        \\
17 & 15.6  & 6.52  & 16.4  & 1.86 & 0.43 & 16.93 & 0.32 & 0.67 &  0.60 &8312-6102   & $109.55\pm1.98$ & $43.72^{+9.70}_{-11.93}$     \\
18 & 17.45 & 3.48  & 18.63 & 2.40 & 0.16 & 18.76 & 0.70 & 0.92 &  1.14 &$-$         & $-$             & $-$                        \\
19 & 15.08 & 8.68  & 16.85 & 7.75 & 0.3  & 15.6  & 0.86 & 1.59 &  1.96 &9869-9102   & $32.79\pm14.05$ & $4.51^{+1.50}_{-0.66}$       \\
20 & 13.98 & 14.73 & 16.68 & 6.34 & 0.2  & 16.23 & 0.78 & 1.76 &  0.85 &8602-12701  & $156.58\pm1.80$ & $14.18^{+2.66}_{-2.63}$      \\
21 & 15.26 & 5.58  & 16.81 & 2.82 & 0.6  & 16.66 & 0.60 & 1.23 &  1.37 &8312-3704   & $66.49\pm7.03$  & $-$   \\
22 & 15.3  & 5.06  & 16.91 & 2.11 & 0.7  & 16.47 & 0.51 & 0.93 &  1.05 &8312-3701   & $73.33\pm4.14$  & $-$    \\
23 & 14.42 & 9.91  & 17.09 & 4.06 & 0.71 & 16.89 & 0.54 & 1.18 &  0.81 &8550-12704  & $165.41\pm3.60$ & $8.46^{+3.37}_{-7.38}$       \\
24 & 14.7  & 12.71 & 15.99 & 4.80 & 0.43 & 15.65 & 1.04 & 1.69 &  0.87 &8604-9101   & $66.67\pm1.71$  & $14.80^{+17.35}_{-11.28}$    \\
25 & 15.57 & 6.39  & 19.26 & 4.88 & 0.33 & 18.76 & 0.86 & 3.43 &  0.99 &$-$         & $-$             & $-$                        \\
26 & 17    & 4.79  & 17.76 & 2.54 & 0.51 & 17.52 & 0.54 & 1.1  &  1.30 &$-$         & $-$             & $-$                        \\
27 & 15.55 & 5.90  & 17.58 & 4.17 & 0.16 & 16.91 & 0.60 & 1.15 &  1.58 &$-$         & $-$             & $-$                        \\
28 & 15.07 & 9.56  & 16.3  & 2.27 & 0.5  & 16.53 & 0.48 & 1.42 &  0.54 &12674-6101  & $168.29\pm2.43$ & $-$     \\
29 & 15.74 & 6.62  & 17.47 & 1.14 & 0.39 & 17.75 & 0.28 & 0.25 &  0.49 &12673-12702 & $54.05\pm4.50$  & $-$    \\
30 & 16.97 & 1.98  & 18.33 & 1.63 & 0.36 & 17.37 & 0.32 & 0.97 &  1.97 &8603-1901   & $88.83\pm6.31$  & $-$ \\
31 & 17.41 & 4.61  & 18.24 & 1.85 & 0.68 & 18.88 & 0.33 & 0.4  &  0.89 &8601-3701   & $24.32\pm6.22$  & $-$     \\
32 & 17.85 & 1.57  & 19.8  & 1.23 & 0.25 & 19.71 & 0.41 & 0.54 &  1.18 &$-$         & $-$             & $-$                        \\
33 & 16.49 & 3.51  & 17.41 & 1.59 & 1.16 & 17.93 & 0.26 & 0.64 &  1.29 &$-$         & $-$             & $-$                        \\
34 & 16.3  & 3.55  & 19.03 & 1.44 & 0.22 & 18.08 & 0.31 & 0.57 &  1.00 &$-$         & $-$             & $-$                        \\
35 & 15.71 & 5.23  & 18.31 & 4.10 & 0.1  & 17.56 & 0.40 & 1.03 &  1.41 &12673-9102  & $36.58\pm5.05$  & $12.61^{+8.56}_{-5.72}$      \\
36 & 14.07 & 7.49  & 16.8  & 2.45 & 0.07 & 17.06 & 0.46 & 0.67 &  0.60 &12674-12703 & $156.58\pm1.53$ & $-$     \\
37 & 15.48 & 6.72  & 17.66 & 3.57 & 0.63 & 17.88 & 0.51 & 1.46 &  1.14 &8602-12703  & $46.13\pm2.70$  & $-$       \\
38 & 15.7  & 6.14  & 16.64 & 2.65 & 0.41 & 16.17 & 0.67 & 1.84 &  1.06 &8312-1902   & $146.85\pm5.05$ & $24.01^{+25.10}_{-16.98}$    \\
39 & 15.56 & 6.12  & 18.91 & 1.33 & 0.27 & 18.13 & 0.26 & 1.42 &  1.69 &8312-12705  & $100.90\pm6.04$ & $6.74^{+3.71}_{-4.19}$ \\
\bottomrule \bottomrule
\end{tabular}
}
\tablefoot{[1] serial number of the barred galaxies in the sample; [2] disc magnitude ($\mathrm{mag}_{\mathrm{disc}}$); [3] disc effective radius ($\mathrm{R}_{\mathrm{e,disc}}$) in kpc; [4] bar magnitude ($\mathrm{mag}_{\mathrm{bar}}$); [5] bar effective radius ($\mathrm{R}_{\mathrm{e,bar}}$) in kpc; [6] bar Sérsic index ($\mathrm{n}_{\mathrm{bar}}$); [7] bulge magnitude ($\mathrm{mag}_{\mathrm{bulge}}$); [8] bulge effective radius ($\mathrm{R}_{\mathrm{e,bulge}}$) in kpc; [9] bulge Sérsic index ($\mathrm{n}_{\mathrm{bulge}}$); [10] normalised projected bar radius with disc scale length ($\mathrm{R}_{\mathrm{bar}}/\mathrm{h}_{\mathrm{disc}}$); [11] MaNGA PlateID; [12] kinematic galaxy position angle ($\mathrm{PA}_{\mathrm{kin}}$) in degrees; [13] bar pattern speed ($\Omega_{\mathrm{bar}}$) in km\,s$^{-1}$\,arcsec$^{-1}$.}
\end{table*}

\begin{figure*}[htb!]
     \centering
     \begin{subfigure}[b]{0.48\textwidth}
         \centering
         \includegraphics[width=\textwidth]{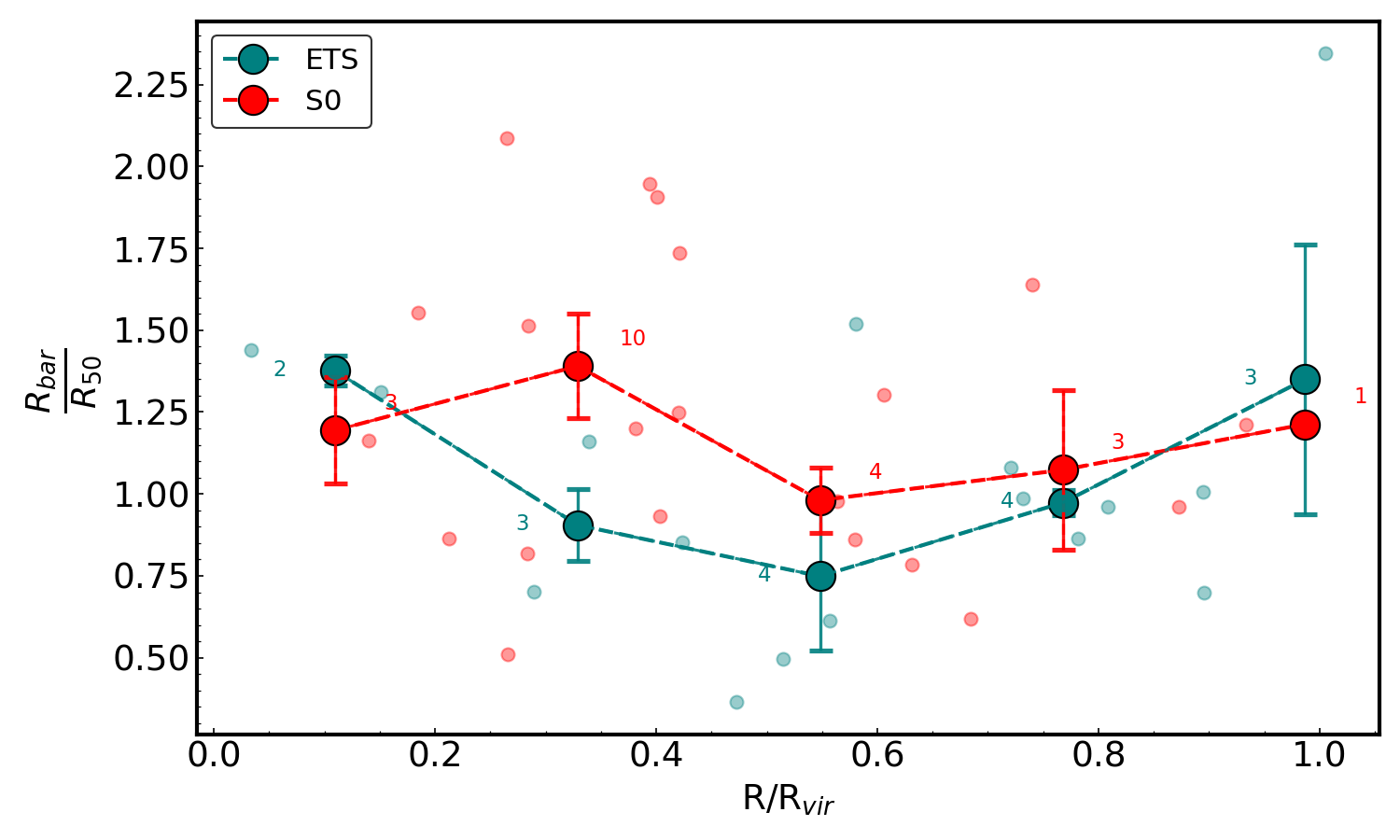}
     \end{subfigure}
     \begin{subfigure}[b]{0.48\textwidth}
         \centering
         \includegraphics[width=\textwidth]{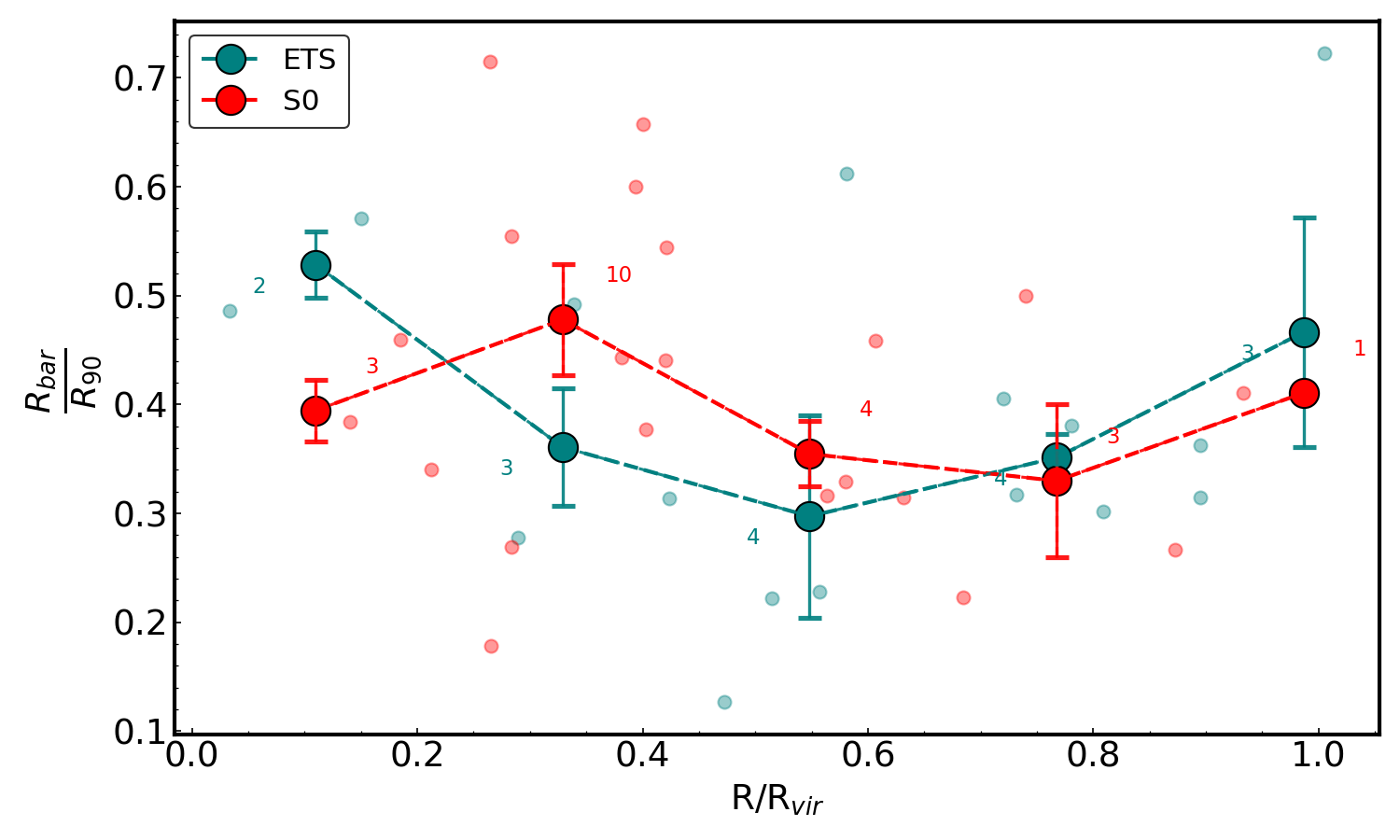}
     \end{subfigure}
     \caption{(left) Normalised bar radius, $R_{\mathrm{bar}}/R_{\mathrm{50}}$, as a function of projected cluster-centric distance for ETS+Bar (teal) and S0+Bar (red) galaxies. (right) Same quantity normalised by $R_{90}$. Here, $R_{50}$ and $R_{90}$ are the radii enclosing 50\% and 90\% of the total galaxy flux, respectively (from the NSA catalogue). Points represent individual galaxies, and solid symbols denote bin-averaged values computed using a radial bin size of 0.5\,Mpc. The numbers shown on the solid symbol indicate the number of galaxies per bin.}
    \label{fig:bar_parameters}
\end{figure*}

\begin{figure*}[ht!]
     \centering
     \begin{subfigure}[b]{0.49\textwidth}
         \centering
         \includegraphics[width=\textwidth]{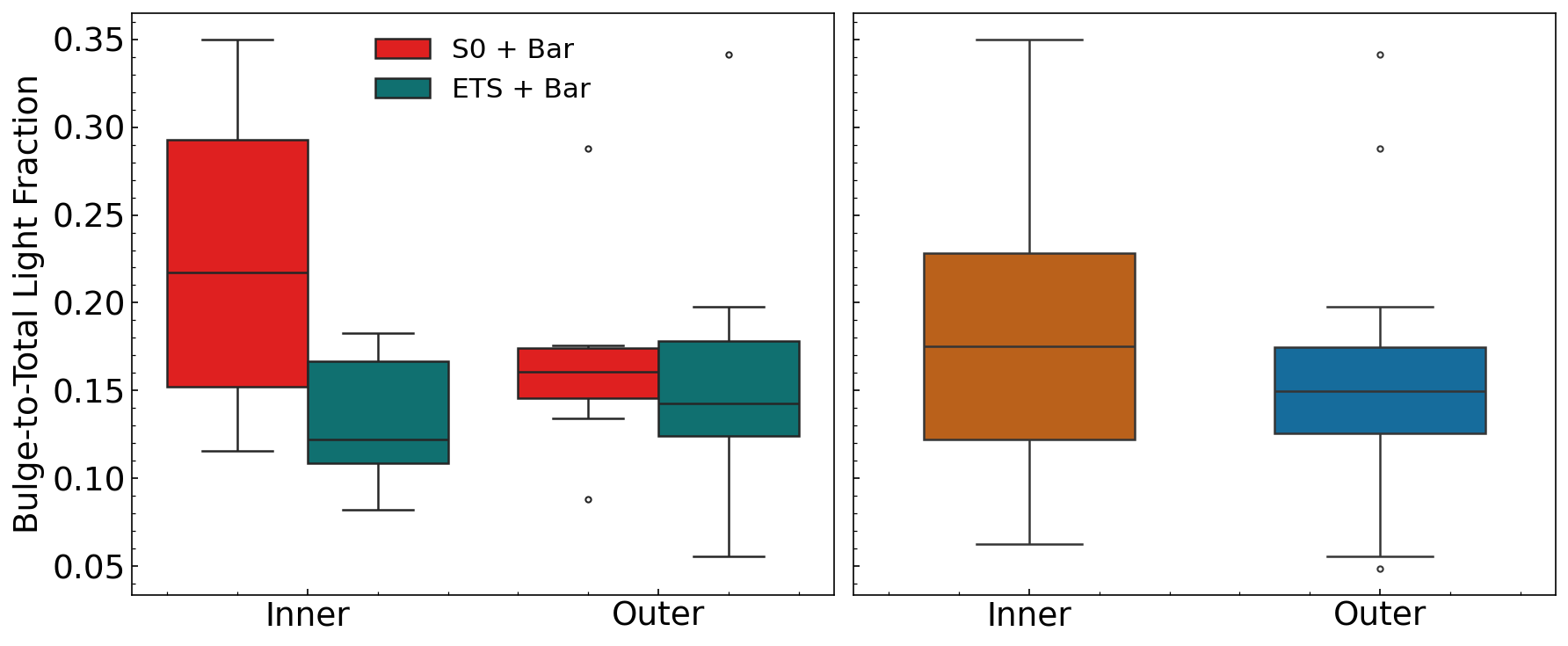}
     \end{subfigure}
     \begin{subfigure}[b]{0.49\textwidth}
         \centering
         \includegraphics[width=\textwidth]{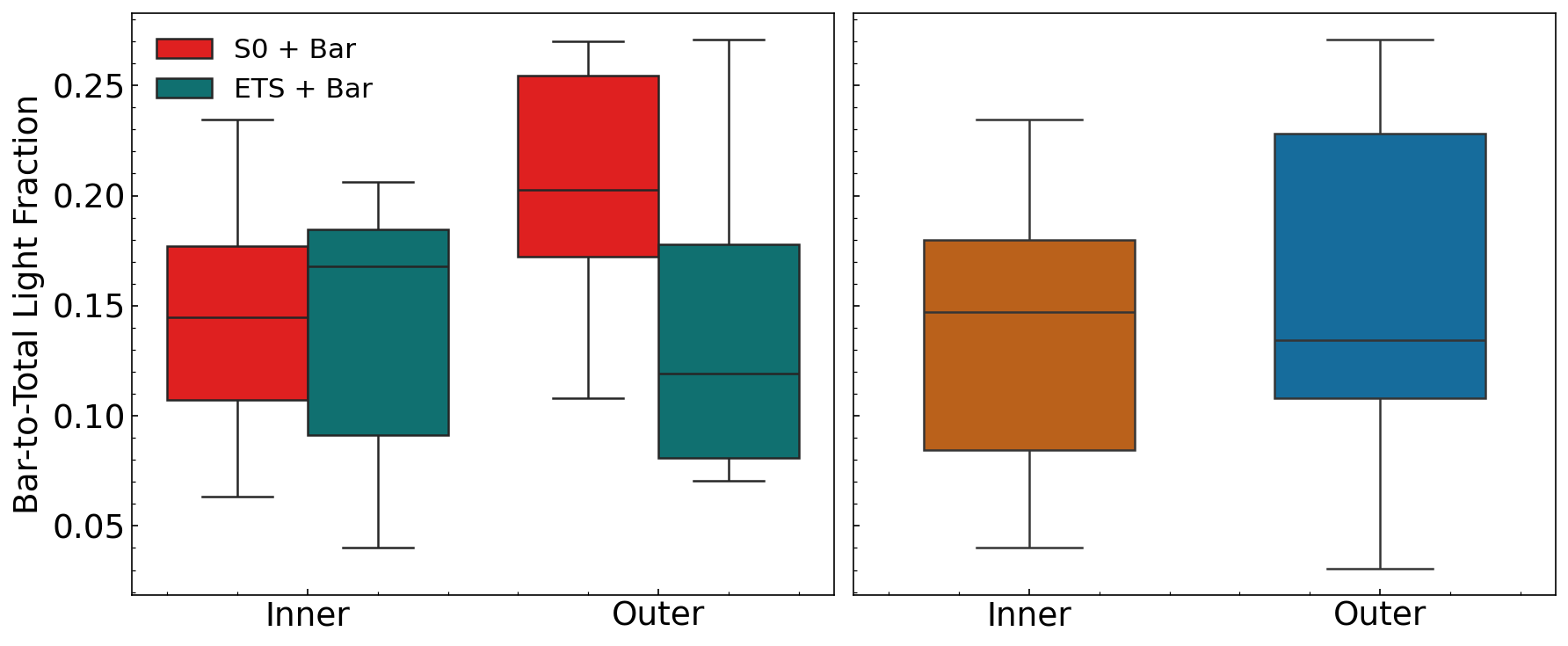}
     \end{subfigure}
     \caption{Box plots showing the distribution of structural parameters for barred galaxies in the inner and outer cluster regions. The two left panels show the bulge-to-total light ratio (B/T), and the two right panels show the bar-to-total light ratio. The red and teal box plots correspond to morphology-separated barred galaxy samples, while the orange and blue box plots show the combined barred galaxy sample. In each box plot, the central line indicates the median, the box spans the interquartile range, and the whiskers represent the full data range of the non-outlier data.}
        \label{fig:box_plot}
\end{figure*}

\begin{table*}[ht!]
\centering
\caption{Median structural and photometric parameters for barred galaxies in the inner and outer regions.} 
\label{tab:inner_outer_all}
\begin{tabular}{lcccccc}
\toprule \toprule
Parameter
& \multicolumn{3}{c}{Inner}
& \multicolumn{3}{c}{Outer} \\
\cmidrule(lr){2-4}\cmidrule(lr){5-7}
& All & ETS+Bar & S0+Bar & All & ETS+Bar & S0+Bar \\
\hline
$\mathrm{n}_{\mathrm{bulge}}$      & 1.08 & 1.33 & 1.00 & 1.15 & 1.14 & 1.03 \\
$\mathrm{n}_{\mathrm{bar}}$        & 0.38 & 0.24 & 0.39 & 0.43 & 0.35 & 0.57 \\
$\mathrm{R}_{\mathrm{e,bar}}$ (kpc)   & 2.41 & 2.41 & 2.11 & 2.65 & 3.27 & 2.06 \\
$\mathrm{R}_{\mathrm{e,bulge}}$ (kpc) & 0.57 & 0.68 & 0.51 & 0.51 & 0.54 & 0.45 \\
$\mathrm{h}_{\mathrm{disc}}$ (kpc) & 3.05 & 3.44 & 2.85 & 3.89 & 4.61 & 2.69 \\
$\mathrm{B/T}$                    & 0.18 & 0.12 & 0.22 & 0.15 & 0.14 & 0.16 \\
$\mathrm{Bar}/T$         & 0.15 & 0.17 & 0.15 & 0.13 & 0.12 & 0.20 \\
$\mathrm{R}_\mathrm{bar}$  (kpc)       & 3.30 & 3.70 & 3.16 & 3.17 & 4.06 & 2.59 \\
$\mathrm{R}_{\mathrm{bar}}/\mathrm{h}_{\mathrm{disc}}$ & 1.05 & 1.09 & 1.06 & 1.01 & 0.97 & 1.16 \\
$\mathrm{R}_{\mathrm{bar}}/\mathrm{R}_{50}$ & 1.18 & 1.01 & 1.25 & 0.96 & 0.97 & 0.97 \\
$\mathrm{R}_{\mathrm{bar}}/\mathrm{R}_{90}$ & 0.44 & 0.40 & 0.44 & 0.33 & 0.34 & 0.32 \\
$\lambda_{\mathrm{R_e}}$ & 0.70 & 0.79 & 0.56 & 0.87 & 0.87 & 0.85 \\
$\mathrm{D4000}_{\mathrm{R_e}}$ & 1.72 & 1.69 & 1.82 & 1.54 & 1.53 & 1.65 \\
\bottomrule \bottomrule
\end{tabular}
\tablefoot{The results are listed for all barred galaxies combined, and separately for ETS+Bar and S0+Bar.}
\end{table*}

\end{appendix}
\end{document}